\documentclass[draftcls,onecolumn]{IEEEtran}
\usepackage{latexsym}
\usepackage{amssymb, amsthm}
\usepackage{graphicx}
\usepackage{mathtools}
\usepackage{amsmath}
\usepackage{algorithm}
\usepackage{algorithmic}
\usepackage{lipsum}
\usepackage{setspace}
\usepackage[caption=false,font=footnotesize]{subfig}
\usepackage{comment}
\usepackage{here}

\newtheorem{theorem}{Theorem}
\newtheorem{corollary}{Corollary}
\newtheorem{example}{Example}
\newtheorem{lemma}{Lemma}
\makeatletter
\newenvironment{myproof}[1][\proofname]{\par
  \normalfont
  \topsep6\p@\@plus6\p@ \trivlist
  \item[\hskip\labelsep{\bfseries #1}\@addpunct{\bfseries.}]\ignorespaces
}{%
  \endtrivlist
}
\renewcommand{\proofname}{\bf Proof}
\makeatother

\newtheorem{remark}{Remark}
\newtheorem{definition}{Definition}

\newtheorem{procedure}{Procedure}
\def\qed{\hfill $\Box$}
\usepackage{color}

\begin{document}
\title{Worst-case Redundancy of Optimal \\ Binary AIFV Codes and their Extended Codes}

\author{Weihua~Hu,
        Hirosuke~Yamamoto,~\IEEEmembership{Fellow,~IEEE,}
        and~Junya~Honda,~\IEEEmembership{Member,~IEEE}% <-this % stops a space
\thanks{W. Hu is with the Department of Computer Science, The University of Tokyo, Bunkyo-ku, Tokyo 113-8656, Japan.\par
E-mail: hu@ms.k.u-tokyo.ac.jp 
\par
H.Yamamoto and J. Honda are with the Department of Complexity Science and Engineering, The University of Tokyo, Kashiwa-shi, Chiba 277-8561, Japan \par E-mail: Hirosuke@ieee.org, honda@it.k.u-tokyo.ac.jp}
\thanks{The material in this paper is presented in part at the IEEE International Symposium on Information Theory, Barcelona, Spain, July 2016.}}
\maketitle

\begin{abstract}
Binary AIFV codes are lossless codes that generalize the class of instantaneous FV codes. The code uses two code trees and assigns source symbols to incomplete internal nodes as well as to leaves. AIFV codes are empirically shown to attain better compression ratio than Huffman codes.
Nevertheless, an upper bound on the redundancy of optimal binary AIFV codes is only known to be 1, which is the same as the bound of Huffman codes. In this paper, the upper bound is improved to 1/2, which is shown to coincide with the worst-case redundancy of the codes.
Along with this, the worst-case redundancy is derived in terms of $p_{\max}\geq$1/2, where $p_{\max}$ is the probability of the most likely source symbol. Additionally, we propose an extension of binary AIFV codes, which use $m$ code trees and allow at most $m$-bit decoding delay. 
We show that the worst-case redundancy of the extended binary AIFV codes is $1/m$ for $m \leq 4.$
\end{abstract}


\IEEEpeerreviewmaketitle

\section{Introduction}
Fixed-to-Variable length (FV) codes map source symbols to variable length codewords, and can be represented by code trees. In the case of a binary instantaneous FV code, source symbols are assigned to leaves of the binary tree. The codeword for each source symbol is then given by the path from the root to the corresponding leaf. 
It is well known by Kraft and McMillan theorems \cite{kraft49}\cite{mcmillan56} that the codeword lengths of any uniquely decodable FV code must satisfy Kraft's inequality, and such codeword lengths can be realized by an instantaneous FV code. Hence, the Huffman code \cite{huffman52}, which can attain the best compression ratio in the class of instantaneous FV codes, is also the optimal code in the class of uniquely decodable FV codes. However, it was implicitly assumed in \cite{mcmillan56} that a single code tree is used for a uniquely decodable FV code. Hence, if we use multiple code trees for a uniquely decodable FV code, it may be possible to attain better compression ratio than Huffman codes.

Recently, Almost Instantaneous Fixed-to-Variable length (AIFV) codes were proposed as a new class of uniquely decodable codes that generalize the class of instantaneous FV codes \cite{yamamoto13}\cite{yamamoto15}. Unlike an instantaneous FV code, which uses only one code tree, an AIFV code is allowed to use multiple code trees. Furthermore, source symbols on the AIFV code trees are assigned to incomplete internal nodes as well as to leaves. In the case of a binary AIFV code \cite{yamamoto15}, two code trees are used in such a way that decoding delay is at most two bits, which is why the code is called {\em almost} instantaneous.

Binary AIFV codes are empirically shown to be powerful in data compression. Not only do the codes attain better compression ratio than Huffman codes, experiments suggest that for some sources, AIFV codes can even beat Huffman codes constructed for ${\cal X}^2$, where ${\cal X}$ is the source alphabet \cite{yamamoto15}. Nonetheless, few theoretical results are known about the codes. In particular, an upper bound on the redundancy (the expected code length minus entropy) of binary AIFV codes is only known to be 1, a trivial bound derived from the fact that binary AIFV codes include Huffman codes. Also, it is conjectured in \cite{yamamoto15} that binary AIFV codes might be able to attain a better compression performance, when more code trees are allowed to be used. %It is an open question to derive a tight upper bound on the redundancy of optimal binary AIFV codes \cite{yamamoto15}.

The main contribution of this paper is two-fold. First, we present a non-trivial theoretical result on the redundancy of optimal binary AIFV codes, suggesting superiority of the codes over Huffman codes. In particular, we show that the worst-case redundancy of optimal binary AIFV codes is $\frac{1}{2}$, which is the same as that of Huffman codes for $\mathcal{X}^2$. Note that for $K=|{\cal X}|$, the size of memory required to store code trees is ${\cal O}(K)$ for a binary AIFV code, while ${\cal O}(K^2)$ for a Huffman code for ${\cal X}^2$ \cite{yamamoto15}. 
\textcolor{black}{Hence, binary AIFV codes require much less memory to store the code trees compared with Huffman codes for ${\cal X}^2$, while attaining comparable compression performance both empirically and theoretically.} 
We also derive the worst-case redundancy of optimal binary AIFV codes in terms of $p_{\max} \geq \frac{1}{2}$, where $p_{\max}$ is the probability of the most likely source symbol. We compare this with its Huffman counterpart \cite{gallager78} and \textcolor{black}{show that for every $p_{\rm max} \geq \frac{1}{2}$, optimal binary AIFV codes improve the worst-case redundancy.} 

Second, we extend the original binary AIFV codes by allowing the code to use more code trees. We show that the extended code with $m$ code trees has at most $m$-bit decoding delay and the worst-case redundancy of the optimal codes is $\frac{1}{m}$ for $m \leq 4.$ Note that the worst-case redundancy of Huffman codes for $\mathcal{X}^m$ per source symbol is also $\frac{1}{m}$. However, the size of a Huffman code tree for $\mathcal{X}^m$ scales exponentially with $m$, while the size of extended AIFV code trees scales only linearly with $m$. \textcolor{black}{Therefore, our extended AIFV codes are much more memory-efficient than Huffman codes for $\mathcal{X}^m$ with the same worst-case redundancy as Huffman codes for $\mathcal{X}^m$, $m \leq 4$.} Our conjecture is that when the size of alphabet is large enough with respect to $m$, the worst-case redundancy of extended AIFV codes is also $\frac{1}{m}$ for $m \geq 5.$ 

The rest of the paper is organized as follows. In Section \ref{preliminaries}, we introduce binary AIFV codes with two code trees \cite{yamamoto15} and some properties of Huffman codes. We extend the binary AIFV codes in Section \ref{extension} so that the codes use more than two code trees. In Section \ref{redundancy}, we give several theorems, \textcolor{black}{which provide the worst-case redundancy} of optimal binary AIFV codes for both the original and extended cases. All the theorems in Sections \ref{redundancy} are proved in Section \ref{proofs}. Section \ref{conclusion} concludes the paper.

\section{Preliminaries} \label{preliminaries}
\subsection{Binary AIFV codes} \label{binary_aifv}
In this section, we introduce the definition of a binary AIFV code proposed in \cite{yamamoto15}.
A binary AIFV code uses two binary code trees, denoted by $T_0$ and $T_1$, in such a way that the code is uniquely decodable and the decoding delay is at most two bits. \textcolor{black}{As conventional FV codes, a source symbol is assigned to a node of the code trees and a code symbol is assigned to an edge of the trees. We begin with a list of properties satisfied by the trees of a binary AIFV code.}

\begin{enumerate}
\item Incomplete internal nodes (nodes with one child) are divided into two categories, master nodes and slave nodes. \label{property-1}
\item Source symbols are assigned to either master nodes or leaves. \label{property-3}
\item The child of a master node must be a slave node, and the master node is connected to its grandchild by code symbols `00'. \label{property-2}
\item The root of $T_1$ has two children. The child connected by `0' from the root is a slave node. The slave node is connected by `1' to its child. \label{property-4}
\end{enumerate}

%Note that in a Huffman code tree, source symbols have to be assigned to leaves in order for the code to be uniquely decodable. 
\begin{remark}
We see from properties \ref{property-1}) and \ref{property-3}) that a binary AIFV code allows source symbols to be assigned to incomplete internal nodes called master nodes. \textcolor{black}{Properties \ref{property-1})--\ref{property-4}) are constraints on the code tree structures, which, combined with encoding and decoding procedures, ensure that the codes are uniquely decodable and decoding delay is at most two bits \cite{yamamoto15}.} 
\end{remark}

\begin{figure}[t]
  \centering
  \includegraphics[width=6.0cm]{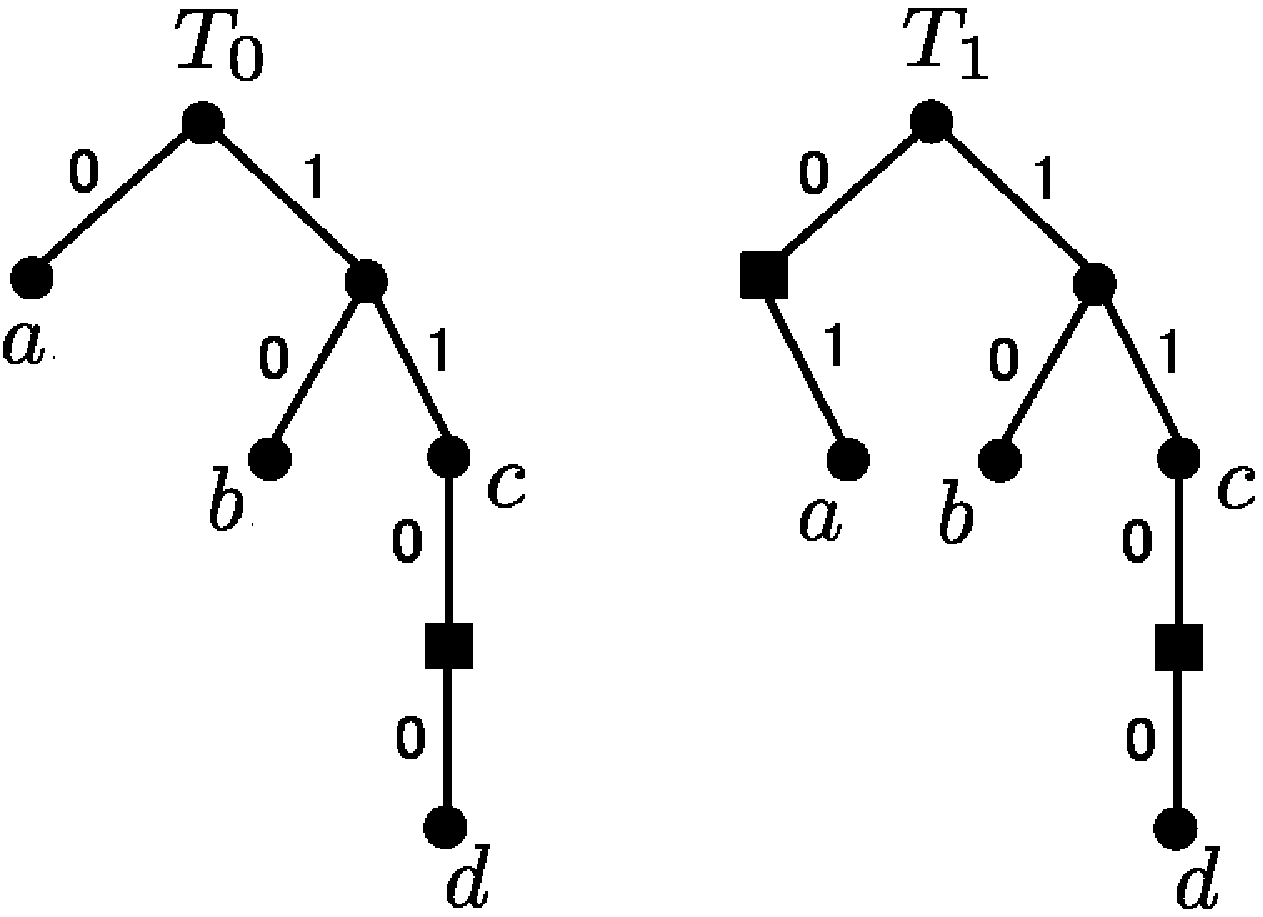}
  \caption{An example of binary AIFV code trees.}
  \label{fig:aifv_example_no_root}
\end{figure}

Fig.~\ref{fig:aifv_example_no_root} illustrates an example of a binary AIFV code for $\mathcal{X} = \{ a,b,c,d\}$, where slave nodes are marked with squares. It is easy to see that the trees satisfy all the properties of a binary AIFV code.

Given a source sequence $x_1 x_2 x_3 \cdots$, an encoding procedure of a binary AIFV code goes as follows. 
\begin{procedure}[Encoding of a binary AIFV code]\label{enc_2aifv}~
\begin{enumerate}
\item Use $T_0$ to encode the initial source symbol $x_1$.
\item When $x_i$ is encoded by a leaf (resp. a master node), then use $T_0$ (resp. $T_1$) to encode the next symbol $x_{i+1}$. \label{enc-2}
\end{enumerate}
\end{procedure}
Using a binary AIFV code of Fig.~\ref{fig:aifv_example_no_root}, a source sequence `$acdbaca$' is encoded to `0.11.1100.10.0.11.01', where dots `.' are inserted for the sake of human readability, but they are not necessary in the actual codeword sequences. The code trees are visited in the order of $T_0 \rightarrow T_0 \rightarrow T_1 \rightarrow T_0 \rightarrow T_0 \rightarrow T_0 \rightarrow T_1$. 

A codeword sequence $y_1 y_2 y_3 \cdots \in \{0,1\}^*$ is decoded as follows.
\begin{procedure}[Decoding of a binary AIFV code]\label{dec_2aifv}~
\begin{enumerate}
\item Use $T_0$ to decode the initial source symbol $x_1$.
\item Trace the codeword sequence as long as possible from the root in the current code tree. Then, output the source symbol assigned to the reached master node or leaf. 
\item If the reached node is a leaf (resp. a master node), then use $T_0$ (resp. $T_1$) to decode the next source symbol from the current position on the codeword sequence.
\end{enumerate}
\end{procedure}
The decoding procedure is guaranteed to visit the code trees in the same order as the corresponding encoding process does \cite{yamamoto15}. The codeword sequence `01111001001101' is indeed decoded to the original source sequence `$acdbaca$', using a sequence of trees in the order of $T_0 \rightarrow T_0 \rightarrow T_1 \rightarrow T_0 \rightarrow T_0 \rightarrow T_0 \rightarrow T_1$.
When all source symbols are assigned to leaves of $T_0$, a binary AIFV code reduces to an instantaneous FV code.  %Therefore, the use of master nodes is essential in binary AIFV codes. 

The following defines the average code length of a binary AIFV code, denoted by $L_{\rm AIFV}$.
\begin{align} \label{ave_len}
L_{\rm AIFV} = P(T_0)L_{T_0} + P(T_1)L_{T_1},
\end{align}
where $P(T_0)$ (resp. $P(T_1)$) is a stationary probability of $T_0$ (resp. $T_1$), and $L_{T_0}$ (resp. $L_{T_1}$) is the average code length of $T_0$ (resp. $T_1$). Consider for instance, a source $\mathcal{X} = \{a, b,c, d \}$ with probabilities $P(a) = 0.45,\ P(b) = 0.3,\ P(c) = 0.2,\ P(d) = 0.05$. Fig.~\ref{fig:aifv_example_no_root} depicts an example of binary AIFV code trees for this source. It follows from the encoding procedure \ref{enc_2aifv} that the transition probabilities $P(T_1|T_0)$ and $P(T_0|T_1)$ are given by $0.2$ and $0.8$, respectively. Therefore, the stationary probabilities $P(T_0)$ and $P(T_1)$ are calculated as 0.8 and 0.2, respectively. Thus, we get $L_{{\rm AIFV}} = 1.65 \cdot 0.8  + 2.1 \cdot 0.2 = 1.74$. On the other hand, it is easy to see that the average code length of the corresponding Huffman code is 1.8. In this example, we see that the binary AIFV code outperforms the Huffman code in terms of compression ratio.

\begin{figure}[t] 
  \centering
  \includegraphics[width=5.0cm]{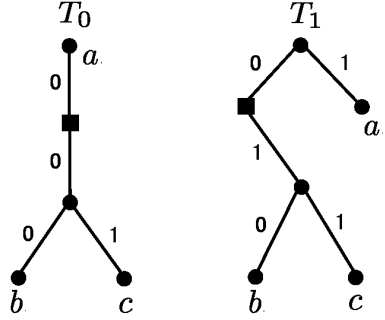}
  \caption{An example of binary AIFV code trees with an source symbol assigned to the root.}
  \label{fig:aifv_example}
\end{figure}

It is worth noting that the root of $T_0$ is allowed to be a master node, to which a source symbol is assigned. Fig.~\ref{fig:aifv_example} illustrates an example of a binary AIFV code for $\mathcal{X} = \{ a,b,c\}$, where $a$ is assigned to the root of $T_0$. The code trees again satisfy all the properties of a binary AIFV code. Using the binary AIFV code of Fig.~\ref{fig:aifv_example}, a source sequence `$aabac$' is encoded to `$\lambda$.1.000.$\lambda$.011', where $\lambda$ is the null codeword corresponding to the root of $T_0.$ 
The code trees are visited in the order of $T_0 \rightarrow T_1 \rightarrow T_0 \rightarrow T_0 \rightarrow T_1$.
\textcolor{black}{We see that in binary AIFV codes, source symbols are sometimes encoded with $\lambda$, i.e., via pure tree transitions without outputting any code symbol.}
It is easy to confirm that the codeword sequence `1000011' is decoded to the original source sequence and decoding procedure visits the code trees in the same order as the encoding procedure does. Note that since the first codeword `1' on the code sequence cannot be traced on $T_0$ of Fig.~\ref{fig:aifv_example}, we output `$a$', which is assigned to the root of $T_0$. 

%Consider for instance, a source $\mathcal{X} = \{a, b,c \}$ with probabilities $P(a) = 0.9$ and $P(b) = P(c) = 0.05$. Let Fig.~\ref{fig:aifv_example} be the corresponding binary AIFV code trees. It follows from the encoding procedure that the transition probabilities $P(T_1|T_0)$ and $P(T_0|T_1)$ are given by $0.9$ and $1.0$. Therefore, the stationary probabilities $P(T_0)$ and $P(T_1)$ are calculated as $\frac{9}{19}$ and $\frac{10}{19}$, respectively. Thus, we get $L_{\rm AIFV} = \frac{9}{19} \cdot 0.3 + \frac{10}{19} \cdot 1.2 \approx 0.774$. It is interesting to see that the average code length of an AIFV code can be less than 1, which is impossible in the case of a Huffman code.

Lastly, we define the redundancy of optimal binary AIFV codes. Let $L_{{\rm OPT}}$ be the average code length of the optimal binary AIFV code for a given source. Then the redundancy of optimal binary AIFV codes denoted by $r_{{\rm AIFV}}$ is defined as
\begin{align} \label{def:aifv_redundancy}
r_{\rm AIFV} \equiv L_{\rm OPT} - H(X),
\end{align}
where $X$ is a random variable corresponding to the source and $H(X)$ is the source entropy. It is shown in \cite{yamamoto15} how we can obtain the optimal binary AIFV code trees for a given source.

\subsection{Sibling property of Huffman codes} \label{sib-sec}
Sibling property was first introduced in $\cite{gallager78}$ as a structural characterization of Huffman codes. Consider a $K$-ary source and let $T_{\rm H}$ denote the corresponding Huffman tree. Let the weight of a leaf be the probability of corresponding source symbol. Also, let the weight of an internal node be defined recursively as the sum of the weights of the children. There are $2K-2$ nodes (except the root) on $T_{\rm H}$. Let $q_1, q_2, \dots, q_{2K-2}$ be the weights of the nodes sorted in a non-increasing order, so that $q_1 \geq q_2 \geq \cdots \geq q_{2K-2}$. By a slight abuse of notation, we identify $q_k$ with the corresponding node itself in the rest of the paper.

We state the sibling property of Huffman codes, which will play an important role in the later proofs of the redundancy of optimal binary AIFV codes.
\begin{definition}[Sibling property]\label{sibling_def}
A binary code tree has the sibling property if there exists a sequence of nodes $q_1, q_2, \dots, q_{2K-2}$, such that for every $k \in \{1, \dots, K-1 \}$, $q_{2k}$ and $q_{2k-1}$ are sibling on the tree.
\end{definition}
\begin{theorem}[{\cite[Theorem 1]{gallager78}}] \label{sib}
A binary instantaneous code is a Huffman code iff the code tree has the sibling property.
\end{theorem}

\subsection{Redundancy upper bounds of Huffman codes}
It is well-known that the worst-case redundancy of Huffman codes is 1. Meanwhile, numerous studies have shown that better bounds on the redundancy can be obtained when a source satisfies some predefined conditions. One such condition concerns with the value of $p_{\max}$ \cite{gallager78}--\cite{johnsen80}, where $p_{\max}$ is the probability of the most likely source symbol. The following is proved by Gallager \cite{gallager78}.
%\cite{gallager78}\cite{manstetten92}\cite{capocelli89}\cite{montgomery87}\cite{johnsen80}

\begin{theorem}[{\cite[Theorem 2]{gallager78}}] \label{huffman_bound}
%For $p_{\max} \geq \frac{1}{2}$, the redundancy of binary Huffman codes is upper bounded by $2 - p_{\max} - h(p_{\max})$, where $h(\cdot)$ is the binary entropy function.
\textcolor{black}{For $p_{\max} \geq \frac{1}{2}$, the worst-case redundancy of binary Huffman codes in terms of $p_{\rm max}$ is given by $2 - p_{\max} - h(p_{\max})$, where $h(\cdot)$ is the binary entropy function.}
\end{theorem}
%Note that the bound provided by Theorem $\ref{huffman_bound}$ is tight in the sense that a source with probabilities $(p_{\max}, 1 - p_{\max} - \delta, \delta)$ satisfies the bound with equality in the limit of $\delta \rightarrow 0$. 
%\textcolor{black}{When an upper bound on the redundancy is tight, the bound is called as the worst-case redundancy.}

\begin{figure}[t]
  \centering
  \includegraphics[width=8cm]{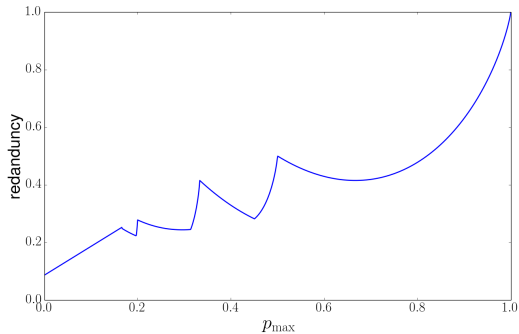}
  \caption{The redundancy upper bounds of Huffman codes in terms of $p_{\max}$.}
  \label{fig:huffman_upper}
\end{figure}

In Fig.~$\ref{fig:huffman_upper}$, we summarize the upper bound results for Huffman codes in terms of $p_{\max}$ \cite{ye02}. \textcolor{black}{For $p_{\max} \geq \frac{1}{6}$, the bound is shown to coincide with the worst-case redundancy of the codes in terms of $p_{\rm max}$.}
We see from Fig.~\ref{fig:huffman_upper} that 
\textcolor{black}{the redundancy upper bound of Huffman codes approaches to its worst-case redundancy, 1, only when source probabilities are extremely biased.} %That is to say, when source probabilities are relatively balanced, the redundancy of a Huffman code can be much smaller than 1.

\section{An extension of binary AIFV codes} \label{extension}
The original binary AIFV code \cite{yamamoto15} in Section \ref{binary_aifv} uses two code trees. In this section, we extend the class of the codes to allow the use of more than two code trees. 
\subsection{Definition of an extended binary AIFV codes}\label{extension2}
We start with redefining a `slave node' and a `master node' that were defined in Section \ref{binary_aifv}.
\begin{definition}[Slave node]\label{slave_node}
A slave node is an incomplete internal node to which no source symbol is assigned. There are two kinds of slave nodes:
\begin{enumerate}
\item A slave-1 node is a slave node which connects to its child by `1.'
\item A slave-0 node is a slave node which connects to its child by `0'. 
\end{enumerate}
\end{definition}
In the later figures, slave nodes are marked by square nodes. To distinguish two kinds of slave nodes, a slave-1 (resp. slave-0) node is marked by a white (resp. black) square node.

\begin{definition}[Master node]\label{master_k}
A master node is defined as a node to which a source symbol is assigned. There are different degrees of master nodes. For $k \geq 1$, a master node of degree $k$ is an incomplete internal node satisfying the following two conditions: 1) $k$ consecutive descendant nodes are slave-0 nodes and 2) the $(k+1)$th descendant is not a slave-0 node. For simplicity, we treat leaves as master nodes of degree 0.
\end{definition}

\begin{figure}[t]
  \centering
  \includegraphics[width=11cm]{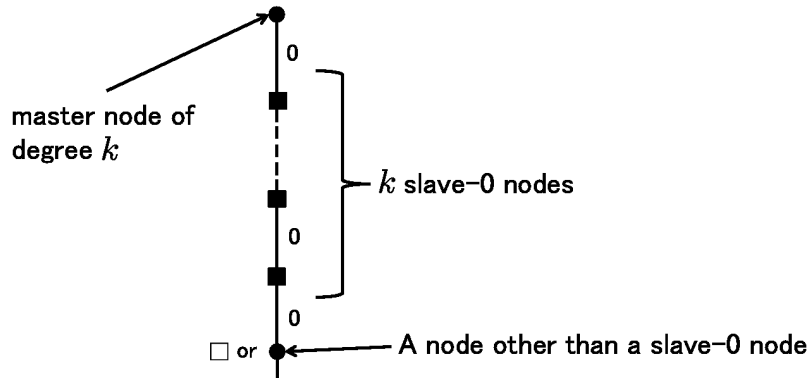}
  \caption{An illustration of \textcolor{black}{a master node of degree $k$.}}
  \label{fig:master_k}
\end{figure}

In contrast to the previous definition of master nodes provided in Section \ref{binary_aifv}, we note that the new definition of master nodes includes leaves. Fig.~\ref{fig:master_k} illustrates a master node of degree $k\geq1$. The master node in Section \ref{binary_aifv} corresponds to a master node of degree 1 in Definition \ref{master_k}. 
%Definition \ref{master_k} implies that a decoding procedure, which traces the codeword sequence on the code trees, has to decode a source symbol assigned to a master node of degree $k \geq 1$ if the subsequent codeword sequence is {\it not} $\underbrace{00\cdots 0}_{k+1}$. 
%We will see that Definition \ref{master_k} is crucial to ensure bounded bits of decoding delay in our extended codes.

Let $m$ be a natural number. As an extension of the original binary AIFV code, we define a {\it binary AIFV-$m$ code}, which uses $m$ code trees and decoding delay is at most $m$ bits. The code trees $T_0,\ T_1,\ \dots, T_{m-1}$ of a binary AIFV-$m$ code are defined by the following properties.

\begin{enumerate}
\item Any node in the code trees is either a slave node, a master node, or a complete internal node. Conforming to Definitions \ref{slave_node} and \ref{master_k}, source symbols are only assigned to master nodes.
\item The degree $k$ of master nodes must satisfy $0 \leq k \leq m-1.$
%\item The root of $T_1$ have two children. The child connected by `0' from the root is a slave-1 node. 
%\item In $T_i$ ($2 \leq i \leq m-1$), a node connected by $\underbrace{00\cdots 0}_{i}$ from the root exists and it is a slave-1 node. 
\textcolor{black}{\item $T_i, \ 1 \leq i \leq m-1$, has a node connected to the root by $\underbrace{00\cdots 0}_{i}$ and it is a slave-1 node.}
%\item In $T_i$ ($2 \leq i \leq m-1$), a node connected by $\underbrace{00\cdots 0}_{j}~(0 \leq j \leq i-2)$ from the root is any node except a master node of degree $k \leq i - j$. A node connected by $\underbrace{00\cdots0}_{i-1}$ from the root is any node except a master node. \label{note}
\end{enumerate}
\begin{remark}
\textcolor{black}
%{It follows from property 3) and Definition \ref{master_k} that the root of $T_i, \ i \geq 2$, can be a master node but its degree should be less than or equal to $i-1$. 
It follows from property 3) and Definition 3 that the root of $T_0$ can be a master node in the same way as the original AIFV codes. Furthermore, the root of $T_i$, $i\geq2$, can also be a master node but its degree should be less than or equal to $i-1$.
\textcolor{black}{As we will see in Lemma \ref{lem:consec}},  when an encoding procedure starts from the root of $T_i,~ 1 \leq i \leq m-1$, the code sequence does {\it not}  start with $\underbrace{00\cdots 0}_{i+1}$. This fact, together with Definition \ref{master_k}, will play a key role in proving unique decodability of binary AIFV codes.
\end{remark}

\begin{remark}
\textcolor{black}{By  property 3) and Definition \ref{master_k}, we see that the root of $T_1$ cannot be a master node and can either be a slave-0 node or a complete internal node. Making the root of $T_1$ to be a slave-0 node (an incomplete internal node without any source symbol) is obviously suboptimal in terms of compression ratio compared to making the root of $T_1$ to be a complete internal node. Hence, in practice, the root of $T_1$ should always be a complete internal node and hence, have two children. Then, for $m=2$, the properties of binary AIFV-$m$ code trees coincide with those of the original binary AIFV code trees explained in Section \ref{binary_aifv}.}
\end{remark}

%\begin{remark}
%In $T_i$, $2 \leq i \leq m-1$, when the node connected by $\underbrace{00\cdots 0}_{j},~0 \leq j \leq i-2,$ from the root is a master node of degree $k$, the $k$ consecutive descendants connected by a sequence of `0' are forced to be slave-0 nodes. This is to meet Definition \ref{master_k} of a master node.
%\end{remark}

We proceed to explain encoding and decoding procedures of the binary AIFV-$m$ codes. Given a source sequence $x_1 x_2 x_3 \cdots$, the encoding procedure is described as follows.

\begin{procedure}[Encoding of \textcolor{black}{a binary AIFV-$m$ code}]~
\begin{enumerate}
\item Use $T_0$ to encode the initial source symbol $x_1$.
\item When $x_i$ is encoded by a master node of degree $k$, then use $T_k$ to encode the next symbol $x_{i+1}$. 
\end{enumerate}
\end{procedure}

A codeword sequence $y_1y_2y_3 \cdots \in \{0,1\}^*$ is decoded as follows.

\begin{procedure}[Decoding of \textcolor{black}{a binary AIFV-$m$ code}]~
\begin{enumerate}
\item Use $T_0$ to decode the initial source symbol $x_1$.
\item Trace the codeword sequence as long as possible from the root in the current code tree. Then, output the source symbol assigned to the reached master node.
\item If the reached node is a master node of degree $k$, then use $T_k$ to decode the next source symbol from the current position on the codeword sequence.
\end{enumerate}
\end{procedure}
In this procedure, the source symbol at a master node of degree $k\geq1$ is decoded after reading subsequent $k+1$ code symbols. Thus, the decoding delay of binary AIFV-$m$ codes is at most $m$ bits.
\textcolor{black}{
\begin{remark}
It follows from the encoding and decoding procedures as well as the properties satisfied by the code trees that binary AIFV codes coincide with binary AIFV-$m$ codes with $m = 2$. Hence, binary AIFV codes are a special case of binary AIFV-$m$ codes.
\end{remark}}

In the following, we show two examples of binary AIFV-$m$ codes to demonstrate how the codes work.

\begin{example} \label{example1}
\upshape
Fig. \ref{fig:3tree} illustrates code trees of a binary AIFV-3 code for $\mathcal{X} = \{ a,b,c, d\}$. We see that $T_0$ has the master node of degree 1, while $T_1$ has the master node of degree 2. 
\begin{figure}[t]
  \centering
  \includegraphics[width=8cm]{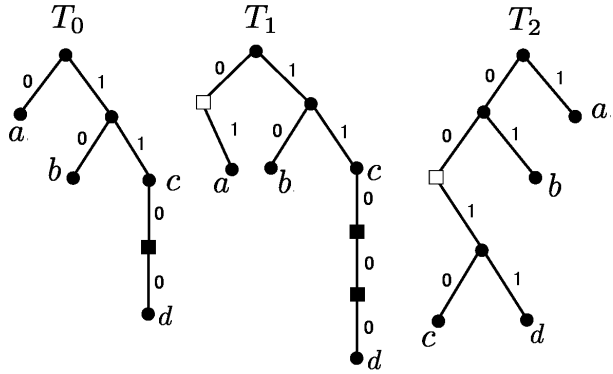}
  \caption{An example of the binary AIFV-3 code trees.}
  \label{fig:3tree}
\end{figure}
Using the binary AIFV code of Fig.~\ref{fig:3tree}, a source sequence `acdccbba' is encoded to `0.11.11000.11.11.01.10.0'. The code trees are visited in the order of $T_0 \rightarrow T_0 \rightarrow T_1 \rightarrow T_0 \rightarrow T_1 \rightarrow T_2 \rightarrow T_0 \rightarrow T_0$. Conversely, it is easy to see that the codeword sequence `01111000111101100' can be decoded to the original source sequence.
\end{example}

The root of $T_i,\ i \neq 1$, can also be a master node. Example~\ref{example2} shows this case. 

\begin{example} \label{example2}
\upshape
Fig.~\ref{fig:3tree2} illustrates a binary AIFV-3 code for $\mathcal{X} = \{ a,b,c\}$, where the roots of $T_0$ and $T_2$ are master nodes of degree 2 and 1, respectively. For instance, a source sequence `$aaabac$' is encoded to `$\lambda . \lambda .1.0000.\lambda.0011$', that is, `100000011'. The code trees are used in the order of $T_0 \rightarrow T_2 \rightarrow T_1 \rightarrow T_0 \rightarrow T_0 \rightarrow T_2$. Conversely, the codeword sequence `100000011' is decoded to the original source sequence.

\begin{figure}[t] 
  \centering
  \includegraphics[width=7cm]{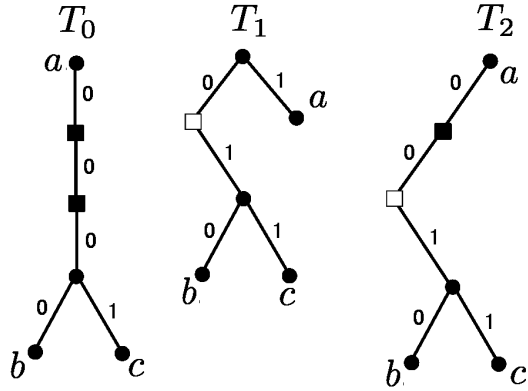}
  \caption{An example of binary AIFV-3 code trees whose roots are master nodes.}
  \label{fig:3tree2}
\end{figure}
\end{example}

We show a lemma, which is crucial to prove the unique decodability of binary AIFV-$m$ codes in Theorem \ref{unique-thm}.

\begin{lemma} \label{lem:consec}
When the encoding procedure starts from the root of $T_i,~ 1 \leq i \leq m-1$, the code sequence does {\it not}  start with $\underbrace{00\cdots 0}_{i+1}$.
\end{lemma}

\begin{myproof}
We prove by induction that for all $2 \leq \ell \leq m$, when the encoding procedure starts from the root of $T_i,~ 1 \leq i \leq \ell - 1$, the code sequence does not start with $\underbrace{00\cdots 0}_{i+1}$.

The base case $(\ell = 2)$ holds since the root of $T_1$ cannot be a master node and any codewords of $T_1$ do not have prefix `00'.

Suppose that the induction hypothesis holds for some $\ell$ such that $2 \leq \ell \leq m-1$. Our goal is to show that the induction hypothesis also holds for $\ell + 1$. To this end, it is sufficient to show that when the encoding procedure starts from the root of $T_{\ell}$, the code sequence does not start with $\underbrace{00\cdots 0}_{\ell+1}$. We consider the following two cases. %for a source symbol encoded by the encoding procedure, when the encoding procedure starts from $T_{\ell}$.
\begin{enumerate}
\item The first symbol in the source sequence is not assigned to a node connected by a sequence of 0's from the root of $T_{\ell}$.
\item The first symbol in the source sequence is assigned to a master node (which we denote as $M$) connected by $\underbrace{00\cdots0}_{j}$, $0\leq j\leq \ell-2$, from the root of $T_{\ell}$.
\end{enumerate}

To begin with, consider the first case. It follows from property 3) of binary AIFV-$m$ code trees that any codewords of $T_{\ell}$ do not have prefix $\underbrace{00\cdots0}_{\ell+1}$. Hence, the encoding procedure would not output a codeword sequence that starts with $\underbrace{00\cdots0}_{\ell+1}$.

We then consider the second case. It follows from Definition \ref{master_k} that the degree of $M$ (which we denote as $k$) must satisfy $1\leq k \leq \ell-j-1$. When the source symbol assigned to $M$ in $T_{\ell}$ is encoded, the encoding procedure would output codeword $\underbrace{00\cdots0}_{j}$ and then use $T_k$ for the next code tree. By the induction hypothesis, when the encoding procedure starts from the root of $T_k$, the codeword sequence does not start with $\underbrace{00\cdots0}_{k+1}$. Therefore, the concatenated codeword sequence does not start with $\underbrace{00\cdots0}_{j}\underbrace{00\cdots0}_{k+1}$, which implies that the sequence does not start with $\underbrace{00\cdots0}_{\ell}$.

In both cases, when the encoding procedure starts from the root of $T_{\ell}$, the code sequence does not start with $\underbrace{00\cdots 0}_{\ell+1}$. \qed
\end{myproof}
%\begin{itemize}
%\item
%For $1\leq i\leq m-1$, any codewords of $T_i$ do not have prefix $\underbrace{00\cdots0}_{i+1}$.
%\item
%In $T_i$, $2\leq i\leq m-1$, when the node connected by $\underbrace{00\cdots0}_{j}$, $0\leq j\leq i-2$, from the root is a master node (which we denote as $M$), its degree $k$ must satisfy $1\leq k \leq i-j-1$ following Definition \ref{master_k}.
%When a source symbol assigned to $M$ in $T_i$ is encoded to codeword $\underbrace{00\cdots0}_{j}$, encoding procedure would use $T_k$ for the next code tree. Since no codewords in $T_k$ have prefix $\underbrace{00\cdots0}_{k+1}$, any concatenated codewords of $T_i$ and $T_k$ do not have prefix $\underbrace{00\cdots0}_{i}$.
%\end{itemize}

\begin{theorem} \label{unique-thm}
Binary AIFV-$m$ codes are uniquely decodable.
\end{theorem}

\begin{myproof}
Let $x_1 x_2 x_3 \cdots$ be a source sequence and let $x'_1 x'_2 x'_3 \cdots$ be the recovered source sequence. Also, let $i_p $ (resp. $j_p$) be a position on codeword sequence $(y_1 y_2 y_3 \cdots)$ from which encoding (resp. decoding) of $x_p$ starts, and let $T^{p}_{{\rm enc}}$ (resp. $T^p_{{\rm dec}}$) be a code tree used to encode (resp. decode) $x_p.$ We prove by induction that $i_p = j_p$ and $T^p_{{\rm enc}} = T^p_{{\rm dec}}$ hold for every $p \geq 1.$

The base case $(p=1)$ clearly holds since $i_1 = j_1=1$ and $T^1_{{\rm enc}} = T^1_{{\rm dec}} = T_0$. 

%We note that $x_0$ is assigned to either a leaf (a master node of degree 0) or a master node of degree $k \geq 1$. If $x_0$ is assigned to a leaf of $T_0$, then the decoding procedure will clearly recover $x_0$. Thus, $x_0 = x'_0$ holds. If $x_n$ is assigned to a master node of degree $k \geq 1$ (which we denote $M$), the encoding procedure will visit $T_k$ for the next symbol $x_{1}$, and thus, the following $k+1$ bits of code sequence will not be $\underbrace{00\cdots 0}_{k+1}.$ Suppose that the decoding procedure does not decode a source symbol at $M$. Then, the decoding procedure should further trace the codeword sequence. However, the only path from the node $M$ starts from $\underbrace{00\cdots 0}_{k+1}$. This is in contradiction with the fact that the following $k+1$ bits of code sequence will not be $\underbrace{00\cdots 0}_{k+1}.$ Therefore, the decoding procedure must decode a source symbol at $M$. Hence, $x_0 = x'_0$ holds. In either case, $x_0 = x'_0$ holds, which establishes the base case.

Suppose that $i_p = j_p$ and $T^p_{{\rm enc}} = T^p_{{\rm dec}}$ hold for a given $p \geq 1$. Our goal is to prove $i_{p+1} = j_{p+1}$ and $T^{p+1}_{{\rm enc}} = T^{p+1}_{{\rm dec}}$. We note that $x_{p}$ is assigned to a master node. First, suppose that $x_{p}$ is assigned to a master node of degree 0, i.e., a leaf. Then, the decoding procedure straightforwardly recovers $x_{p}$ with no decoding delay. Thus, $x'_{p} = x_{p}$ holds and so does $i_{p+1} = j_{p+1}$.  Then, both the encoder and decoder visit $T_0$ for the next source symbol $x_{p+1}$. Hence, $T^{p+1}_{{\rm enc}} = T^{p+1}_{{\rm dec}}$ holds. Next, suppose that $x_{p}$ is assigned to a master node of degree $k \geq 1$ (which we denote as $M$). Then, the encoding procedure visits $T_k$ for the next symbol $x_{p+1}$. %Since the prefix of $T_k$ is not $\underbrace{00\cdots 0}_{k+1},$ the subsequent $k+1$ bits of the code sequence is also not $\underbrace{00\cdots 0}_{k+1}.$ 
\textcolor{black}{We note from Lemma \ref{lem:consec} that the code sequence does not start with $\underbrace{00\cdots0}_{k + 1}$, when the encoding procedure starts from the root of $T_k$.}
On the other hand, since $M$ is a master node of degree $k$, we see from Definition \ref{master_k} that the prefix of all codeword sequences on $T_{{\rm enc}}^p$ starting from $M$ is $\underbrace{00\cdots 0}_{k+1}$. Thus, the decoding procedure cannot trace any codeword starting from $M$. Therefore, the decoding procedure decodes a source symbol at $M$, which leads the decoder to use $T_k$ for the next source symbol $x_{p+1}$. Hence, $x_p= x'_p,\ i_{p+1} = j_{p+1}$ and $T^{p+1}_{{\rm enc}} = T^{p+1}_{{\rm dec}}$ hold. In either cases, $i_{p+1} = j_{p+1}$ and $T^{p+1}_{{\rm enc}} = T^{p+1}_{{\rm dec}}$ hold.

From the above induction, we have that $i_p = j_p$ and $T^p_{{\rm enc}} = T^p_{{\rm dec}}$ and hence $x_p$ =  $x'_p$, which means that binary AIFV-$m$ codes are uniquely decodable.
\qed  
\end{myproof}

We proceed to define an average code length $L_{{\rm AIFV}}$ and redundancy $r_{{\rm AIFV}}$ of binary AIFV-$m$ code.
\begin{definition}\label{ave_general}
Let $\{T_k\}_{k = 0, \dots, m-1}$ be a set of the code trees. Let $\{L_{T_k}\}_{k = 0, \dots, m-1}$ and $\{P(T_k) \}_{k = 0, \dots, m-1} $ be a set of average code lengths and a set stationary probabilities of the code trees, respectively. The average code length of the binary AIFV-$m$ code is defined as follows.
\begin{align} \label{av_len}
L_{{\rm AIFV}} = \sum_{k = 0}^{m-1} P(T_k) L_{T_k}.
\end{align}
\end{definition}

Redundancy of optimal binary AIFV-$m$ codes is defined in the same way as \eqref{def:aifv_redundancy}.

Based on Definition \ref{ave_general}, we show how to calculate the average code length of binary AIFV-$m$ code. For example, consider the code trees in Fig.~\ref{fig:3tree} for a source $\mathcal{X} = \{a,b,c, d\}$ with probabilities $P(a) = 0.65,\ P(b) = 0.2,\ P(c) = 0.1$ and $P(d) = 0.05.$ The transition matrix $R$ is given by
\begin{align}
R = 
\begin{pmatrix}
0.9 & 0.1 & 0 \\
0.9 & 0 & 0.1 \\
1 & 0 & 0
\end{pmatrix},
\end{align}
where $R_{ij}$ denotes the transition probability $P(T_j|T_i)$ from $T_i$ to $T_j$. The vector $\mbox{\boldmath $p$}=(P(T_0),P(T_1),\dots,P(T_{m-1}))^{\top}$ of stationary probabilities satisfies
\begin{align}
\mbox{\boldmath $p$}^{\top} R = \mbox{\boldmath $p$}^{\top}. 
\end{align}
Therefore, $\mbox{\boldmath $p$}$ is the normalized left eigenvector of $R$ with eigenvalue of 1. Hence, we get $P(T_0) =\frac{100}{111} ,\ P(T_1) = \frac{10}{111}$ and $P(T_2) = \frac{1}{111}.$ By Definition \ref{ave_general}, the average code length $L_{{\rm AIFV}}$ is calculated as
\begin{align}
L_{{\rm AIFV}} = \frac{100}{111} \cdot 1.45 + \frac{10}{111} \cdot 2.15 + \frac{1}{111} \cdot 1.65 &= \frac{168.15}{111}  \approx 1.515.
\end{align} 

We proceed to show the calculation of the average code length of the code in Fig.~\ref{fig:3tree2} for a source with probabilities $P(a) = 0.98, \ P(b) = 0.01,$ and $P(c) = 0.01$. The transition matrix $R$ is obtained as
\begin{align} \label{R2}
R = 
\begin{pmatrix}
0.02 & 0 & 0.98 \\
1 & 0 & 0 \\
0.02 & 0.98 & 0
\end{pmatrix}.
\end{align}
%The stationary probabilities corresponds to the normalized left eigenvector of $R$ of Eq.~\eqref{R2} with eigenvalue of 1. 
Hence, we get the stationary probabilities as $P(T_0) = \frac{2500}{7351} \approx 0.340,\ P(T_1) = \frac{2401}{7351}\approx 0.327,$ and $P(T_2) = \frac{2450}{7351} \approx 0.333$. From \eqref{av_len}, the average code length $L_{{\rm AIFV}}$ is calculated as 
\begin{align}
L_{{\rm AIFV}} \approx 0.340 \cdot 0.08 + 0.327 \cdot 1.04 + 0.333 \cdot 0.08 \approx 0.394. 
\end{align} 

Note that average code lengths of $T_0$ and $T_2$ are much smaller than 1. This is because the most likely source symbol `$a$' is assigned to the roots of the trees. Compared with Fig.~\ref{fig:aifv_example}, which uses only two code trees, the binary AIFV-3 code shown in Fig.~\ref{fig:3tree2} uses an additional code tree $T_2$. The introduction of $T_2$ enables the extended code to compress the source better than the original binary AIFV code in Fig.~\ref{fig:aifv_example}. 

\section{\textcolor{black}{Worst-case redundancy of optimal binary AIFV codes and their extended codes}} \label{redundancy} 
In this section, we first show the \textcolor{black}{worst-case redundancy of optimal binary AIFV codes \cite{yamamoto15}, which directly suggests superiority} of optimal binary AIFV codes over Huffman codes in terms of compression ratio. We then show \textcolor{black}{the worst-case redundancy} of optimal binary AIFV-$m$ codes for $m \leq 4$. The result suggests that the binary AIFV-$m$ codes further improve the worst-case redundancy of the original binary AIFV codes. The proofs of the theorems in this section are given in Section \ref{proofs}.
\subsection{\textcolor{black}{Worst-case redundancy of optimal binary AIFV codes}}
\begin{theorem}\label{thm-1}
%For $p_{\max} \geq \frac{1}{2}$, the redundancy of optimal binary AIFV codes is upper bounded by $f(p_{\max}),$
For $p_{\max} \geq \frac{1}{2}$, \textcolor{black}{the worst-case redundancy of optimal binary AIFV codes in terms of $p_{\rm max}$ is given by $f(p_{\max}),$}
where $f(x)$ is defined as follows.
\begin{align}
f(x) = \left\{ \begin{array}{ll}
x^2 - 2x + 2 - h(x) & \text{{\rm if} $\frac{1}{2} \leq x \leq \frac{-1+\sqrt{5}}{2}$,} \\
\frac{-2x^2+x+2}{1+x} - h(x) & \text{{\rm if} $\frac{-1+\sqrt{5}}{2} \leq x < 1$.} \\
\end{array} \right.
\end{align}
%The above bound is tight in the sense that there exists a source distribution for any $\epsilon>0$ such that the redundancy is larger than $f(p_{\max})-\epsilon.$
\end{theorem}

\begin{figure}[t] 
  \centering
  \includegraphics[width=8cm]{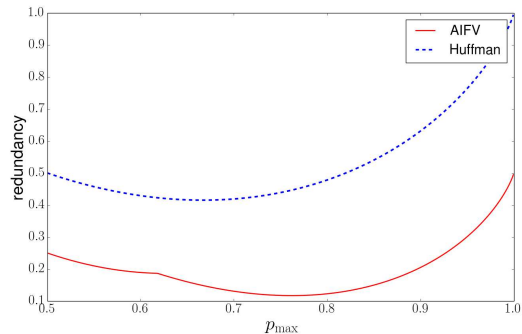}
  \caption{\textcolor{black}{Comparison between optimal binary AIFV codes and Huffman codes in terms of the worst-case redundancy for sources with $p_{\max}$.}}
  \label{fig:comparison}
\end{figure}

\textcolor{black}{Fig.~\ref{fig:comparison} compares the worst-case redundancy given by Theorems \ref{huffman_bound} and \ref{thm-1}}. We see that the worst-case redundancy of optimal binary AIFV codes is smaller than that of Huffman codes for every $p_{\max} \geq \frac{1}{2}$. 

We also get Theorem \ref{thm-2}, covering the case of $p_{\max} < \frac{1}{2}$. 
\begin{theorem} \label{thm-2}
For $p_{\max} < \frac{1}{2}$, the redundancy of optimal binary AIFV codes is at most $\frac{1}{4}.$
\end{theorem}
\textcolor{black}{The bound given by Theorem \ref{thm-2} may not be tight for each $p_{\rm max} \in \left( 0, \frac{1}{2}\right)$ and does not necessarily coincide with the worst-case redundancy of optimal binary AIFV codes for sources with $p_{\rm max} \in \left( 0, \frac{1}{2}\right)$.} Yet, the derived bound is sufficient to prove Corollary \ref{cor-1}, which follows immediately from Theorems \ref{thm-1} and \ref{thm-2}.
\begin{corollary} \label{cor-1}
%The tight upper bound on the redundancy of optimal binary AIFV codes is $\frac{1}{2}$. 
\textcolor{black}{The worst-case redundancy of optimal binary AIFV codes is $\frac{1}{2}$.}
\end{corollary}
\textcolor{black}{Note that the worst-case redundancy given in Corollary \ref{cor-1} is the same as that of Huffman codes constructed for $\mathcal{X}^2$.} 
\textcolor{black}{It is also empirically shown that for some sources}, optimal binary AIFV codes can beat Huffman codes for $\mathcal{X}^2$ \cite{yamamoto15}. Meanwhile, in terms of memory efficiency, a binary AIFV code only requires $\mathcal{O} (K)$ for storing code trees, while a Huffman code for $\mathcal{X}^2$ needs $\mathcal{O}(K^2)$, where $K = |\mathcal{X}|$. 
\textcolor{black}{These suggest} that binary AIFV codes are more memory-efficient than Huffman codes for $\mathcal{X}^2$, while attaining comparable compression performance \textcolor{black}{both empirically and theoretically}.

\subsection{\textcolor{black}{Worst-case redundancy of optimal binary AIFV-$m$ codes}}
Do binary AIFV-$m$ codes provide further compression improvement over the original binary AIFV codes? 
In the following, we obtain \textcolor{black}{the worst-case redundancy of optimal binary AIFV-$m$ codes for $m = 3$ and 4.} 
These results suggest that binary AIFV-$m$ codes can benefit from the use of more code trees. 

%We first provide a lower bound on the worst-case redundancy of optimal uniquely decodable codes whose decoding delay is at most $m$ bits.
%\begin{theorem} \label{lower_bb}
%A tight upper bound on the redundancy of optimal uniquely decodable codes whose decoding delay is at most $m$ bits is larger than or equal to $\frac{1}{m}$.
%\end{theorem}

%Since binary AIFV-$m$ codes are uniquely decodable and the decoding delay is at most $m$ bits, we immediately get Corollary \ref{lower_bb}.
%We first provide a lower bound on the worst-case redundancy of optimal binary AIFV-$m$ codes.   

%It is worth noting that not all uniquely decodable codes whose decoding delay is at most $m$ bits can be represented as binary AIFV-$m$ codes. However, the reassuring fact that Theorems \ref{bound3} and \ref{bound4} suggest is that the worst-case redundancy of optimal binary AIFV-$m$ codes attain the theoretical lower bound of Theorem \ref{lower_bb} for $m = 3$ and 4. That is, for $m = 3$ and 4, the worst-case redundancy of optimal binary AIFV-$m$ is the lowest possible in the class of uniquely decodable codes with at most $m$ bits of decoding delay.

%We now provide the worst-case redundancy of the optimal binary AIFV-$m$ for $m = 3$ and 4. 

\begin{theorem}\label{lower_bb}
%The tight upper bound on the redundancy of binary AIFV-$m$ codes is larger than or equal to $\frac{1}{m}$.
\textcolor{black}{The worst-case redundancy of optimal binary AIFV-$m$ codes is larger than or equal to $\frac{1}{m}$.}
\end{theorem}

\begin{theorem} \label{bound3}
%The tight upper bound on the redundancy of optimal binary AIFV-$3$ codes is $\frac{1}{3}$.
\textcolor{black}{The worst-case redundancy of optimal binary AIFV-$3$ codes is $\frac{1}{3}$.}
\end{theorem}

\begin{theorem} \label{bound4}
%The tight upper bound on the redundancy of optimal binary AIFV-$4$ codes is $\frac{1}{4}$.
\textcolor{black}{The worst-case redundancy of optimal binary AIFV-$4$ codes is $\frac{1}{4}$.}
\end{theorem}

\begin{theorem} \label{pmax_one}
%The redundancy of optimal binary AIFV-$m$ is at most $\frac{1}{m}$ for $p_{\max} \rightarrow 1,$ that is, there exists $\epsilon>0$ such that the redundancy is at most $1/m$ for all $p_{\max} \in (1-\epsilon, 1)$.
\textcolor{black}{The redundancy of optimal binary AIFV-$m$ codes is $\frac{1}{m}$ for $p_{\max} \rightarrow 1.$} %that is, there exists $\epsilon>0$ such that the redundancy is at most $1/m$ for all $p_{\max} \in (1-\epsilon, 1)$.
\end{theorem}

\textcolor{black}{Although the above theorems provide the worst-case redundancy of optimal binary AIFV-$m$ codes for the case of $m = 3$ and 4, 
we conjecture that for any natural number $m$, the worst-case redundancy of optimal AIFV-$m$ codes is also $\frac{1}{m}$.}
Our justification for the conjecture is as follows.
Huffman codes and the optimal AIFV-$m$ codes for $m \leq 4$ have the worst-case redundancy when $p_{\max} \to 1$.
Thus, it is natural to assume that the same property also holds for $m \geq 5$.
If this assumption is true, then the worst-case redundancy is $\frac{1}{m}$ from Theorem \ref{pmax_one},
although it is an open question whether this assumption holds or not.

\section{Proofs of Theorems \ref{thm-1}--\ref{pmax_one}} \label{proofs}
We start with proving Theorems \ref{thm-1} and \ref{thm-2}, which provide upper bounds on the redundancy of optimal binary AIFV codes \textcolor{black}{in terms of $p_{\rm max}$, where $p_{\rm max}$ is the probability of the most likely source symbol.} The key to the proofs is to construct binary AIFV code trees by transforming a Huffman code tree. Then, we can utilize the structural property of the Huffman tree, the sibling property introduced in Section \ref{sib-sec}, in evaluating the redundancy of the binary AIFV code. We first prepare three lemmas, which provide useful inequalities for the later evaluations on the bounds. Note that $q_1, q_2, \cdots, q_{2K-2}$ are defined in Section \ref{sib-sec}.

\begin{lemma} \label{lem-1}
Consider a Huffman tree and suppose that $q_{2k-1}$ is not a leaf. Then, $q_{2k-1} \leq 2q_{2k}$ holds.
\end{lemma}
\begin{myproof}
Let $q'_1$ and $q'_2$ be children of $q_{2k-1}$. In the construction of the Huffman tree, $q'_1$ and $q'_2$ are merged before $q_{2k-1}$ and $q_{2k}$ are merged. Thus, $q'_1 \leq q_{2k}$ and $q'_2 \leq q_{2k}$ hold. Therefore, we get $q_{2k-1} = q'_1 + q'_2 \leq 2q_{2k}.$ \qed
\end{myproof}

\begin{lemma} \label{lem-2}
Assume $0 < 2w_2 < w_1$ and let $q \in \left[0, \frac{1}{2}\right]$ be arbitrary. Then, 
\begin{align}
2w_2 - (w_1 + w_2) & h\left( \frac{w_2}{w_1 + w_2} \right) + q w_1 < q \left( w_1 - w_2 \right). \label{ineq-3}
\end{align}
\end{lemma}
\begin{myproof}
Let $c \equiv \frac{w_1}{w_2}> 2$ and define $g(x) \equiv h(x) - 2x$. Subtracting the LHS of \eqref{ineq-3} from the RHS, we get
\begin{align}
- q&w_2 - 2w_2 + (1+c)w_2 \cdot \left( g \left(\frac{1}{1+c} \right) + \frac{2}{1+c} \right)\nonumber  \\
%&= -q  q_{2k} + (1+c) \cdot q_{2k}  g \left(\frac{1}{1+c} \right) \nonumber \\
%&\geq -\frac{1}{2} q_{2k} + (1+c) \cdot q_{2k}  g \left(\frac{1}{1+c} \right) \nonumber \\
&= w_2 \cdot \left( - q + (1+c)  \cdot g \left(\frac{1}{1+c} \right) \right) \nonumber \\
&\geq w_2 \cdot \left( - \frac{1}{2} + (1+c)  \cdot g \left(\frac{1}{1+c} \right) \right) \nonumber \\
&>0\label{eq:z}.
\end{align}
The last inequality follows from $\inf _{c > 2} (1+c)g(\frac{1}{1+c}) = 0.754 \cdots > \frac{1}{2}$. \qed
\end{myproof}

\begin{lemma} \label{lem-3}
If $q_{2k-1} \leq 2q_{2k}$, then
\begin{align}
(q_{2k-1} + q_{2k}) \left( 1 - h\left( \frac{q_{2k}}{q_{2k-1} + q_{2k}}\right) \right) \leq \frac{1}{4} \left( q_{2k-1} - q_{2k}\right).
\end{align}
\end{lemma}
\begin{myproof}
Since $q_{2k} \leq q_{2k-1} \leq 2q_{2k}$, it follows that $\frac{1}{3} \leq \frac{q_{2k}}{q_{2k-1} + q_{2k}} \leq \frac{1}{2}$. Further, since $\frac{1}{2}x + \frac{3}{4} \leq h(x)$ holds for $\frac{1}{3} \leq x \leq \frac{1}{2}$,
\begin{align}
(q_{2k-1} &+ q_{2k}) \left( 1 - h\left( \frac{q_{2k}}{q_{2k-1} + q_{2k}}\right) \right) \nonumber \\
&\leq (q_{2k-1} + q_{2k}) \left(1 - \frac{q_{2k}}{2(q_{2k-1} + q_{2k})}  - \frac{3}{4} \right) \nonumber \\
&= \frac{1}{4} (q_{2k-1} - q_{2k}). 
\end{align}
\qed
\end{myproof}

\begin{figure}[t] 
  \centering
  \includegraphics[width=7.0cm]{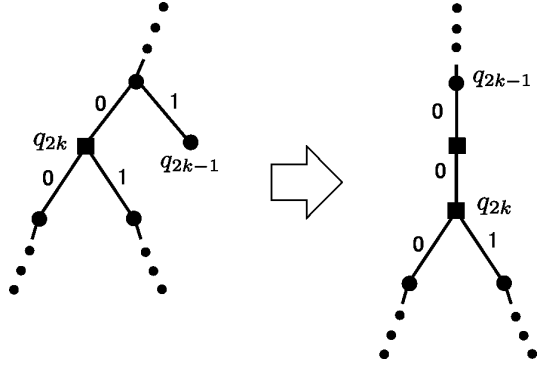}
  \caption{The conversion from a sibling pair ($q_{2k-1},q_{2k}$) into a master node and its grandchild.}
  \label{fig:rotation}
\end{figure}

Now, consider a $K$-ary source and let $T_{\rm H}$ be a Huffman code tree for the source. We transform $T_{\rm H}$ into a new tree $T_{\rm base}$ using Algorithm \ref{alg1}.
\begin{algorithm} [t]                    
\caption{Transformation of $T_{{\rm H}}$ into $T_{{\rm base}}$.}         
\label{alg1}                          
\begin{algorithmic}  
\STATE \FOR{$k = 2, \dots, K-1$} 
\IF{$q_{2k-1}$ is a leaf and $2q_{2k} < q_{2k-1}$}
\STATE Convert the sibling pair ($q_{2k}, \ q_{2k-1}$) into a master node and its grandchild. $\cdots$ ($\ast$)
\ENDIF
\ENDFOR
\end{algorithmic}
\end{algorithm}
The conversion shown by ($\ast$) in Algorithm \ref{alg1} is illustrated in Fig.~\ref{fig:rotation}. It is an operation that lifts up $q_{2k-1}$ to make a master node and pulls down the entire subtree of $q_{2k}$ to one lower level. Two nodes, $q_{2k-1}$ and $q_{2k}$, are then connected by code symbols `00'. 
Let $\mathcal{K}$ be the set of indices whose corresponding sibling pairs are converted by Algorithm \ref{alg1} and let $\mathcal{U}$ denote the set of indices of entire sibling pairs, so that
\begin{align*}
\mathcal{K} &= \{k \in \{ 2, \dots, K-1\}|  q_{2k-1}\text{ is a leaf and } 2q_{2k} < q_{2k-1}\},\\
\mathcal{U} &= \{1,\dots, K-1\}. 
\end{align*}

\begin{lemma} \label{lem-4}
For $\frac{1}{2} \leq q_1 \leq \frac{2}{3}$, the redundancy of optimal binary AIFV codes \textcolor{black}{in terms of $q_1$} is upper bounded by $q_1^2 - 2q_1 + 2 - h(q_1).$
\end{lemma}
\begin{myproof}
Let $T_0$ be $T_{\rm base}$ and transform $T_{\rm base}$ into $T_1$ by the operation described in Fig.~\ref{fig:trans1}.
\begin{figure}[t]
  \centering
  \includegraphics[width=7.0cm]{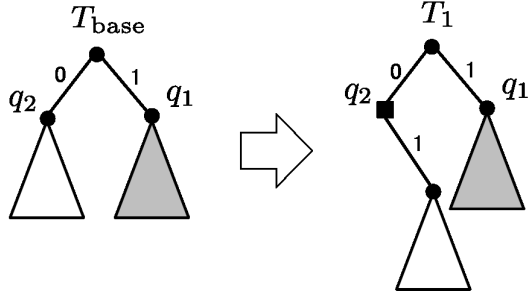}
  \caption{Transformation of $T_{\rm base}$ into $T_1$.}
  \label{fig:trans1}
\end{figure}
It is easy to see that $T_0$ and $T_1$ are valid binary AIFV code trees, satisfying all the properties mentioned in Section \ref{binary_aifv}. 
\begin{spacing}{1.08}
%The total probability assigned to the incomplete internal nodes is {\small $\sum_{k \in \mathcal{K}}q_{2k-1}$} for both {\small $T_0$} and {\small $T_1$}. Thus, it follows from the encoding procedure \ref{enc-2}) in Section \ref{binary_aifv} that the transition probabilities, {\small $P(T_1|T_0)$} and {\small $P(T_0|T_1)$}, are given by {\small $\sum_{k \in \mathcal{K}}q_{2k-1}$} and {\small$1- \sum_{k \in \mathcal{K}}q_{2k-1}$}, respectively. Therefore, the stationary probabilities, {\small $P(T_0)$} and {\small$P(T_1)$}, are calculated as {\small $1- \sum_{k \in \mathcal{K}}q_{2k-1}$} and {\small $\sum_{k \in \mathcal{K}}q_{2k-1}$}, respectively. Then, we have from \eqref{ave_len} and {\small$L_{\rm OPT} \leq L_{\rm AIFV}$} that
The total probability assigned to master nodes is $\sum_{k \in \mathcal{K}}q_{2k-1}$ for both $T_0$ and $T_1$. Thus, it follows from the encoding procedure \ref{enc-2}) in Section \ref{binary_aifv} that the transition probabilities, $P(T_1|T_0)$ and $P(T_0|T_1)$, are given by $\sum_{k \in \mathcal{K}}q_{2k-1}$ and $1- \sum_{k \in \mathcal{K}}q_{2k-1}$, respectively. Therefore, the stationary probabilities, $P(T_0)$ and $P(T_1)$, are calculated as $1- \sum_{k \in \mathcal{K}}q_{2k-1}$ and $\sum_{k \in \mathcal{K}}q_{2k-1}$, respectively. Then, we have from \eqref{ave_len} and $L_{\rm OPT} \leq L_{\rm AIFV}$ that
\end{spacing}

\begin{align}
L_{{\rm OPT}} &\leq L_{T_0}\cdot P(T_0) + L_{T_1} \cdot P(T_1)\nonumber \\
&= \left( \sum_{k \in \mathcal{U}\setminus \mathcal{K}}(q_{2k-1} + q_{2k}) + \sum_{k\in \mathcal{K}} 2q_{2k} \right)  \cdot \left( 1- \sum_{k \in \mathcal{K}}q_{2k-1}\right) \nonumber \\
&\qquad +\left( q_2 + \sum_{k \in \mathcal{U}\setminus \mathcal{K}}(q_{2k-1} + q_{2k}) + \sum_{k\in \mathcal{K}} 2q_{2k} \right)\cdot \sum_{k \in \mathcal{K}}q_{2k-1} \nonumber \\
&= \sum_{k \in \mathcal{U}\setminus \mathcal{K}}(q_{2k-1} + q_{2k}) + \sum_{k\in \mathcal{K}} 2q_{2k}  + q_2\cdot \sum_{k \in \mathcal{K}}q_{2k-1} \nonumber \\
&= \left( q_1 + q_2 \right) + \sum_{k \in \mathcal{K}} \left(2q_{2k} + q_2q_{2k-1} \right) +\sum_{k \in \mathcal{U} \setminus (\mathcal{K} \cup \{ 1\})} (q_{2k-1} + q_{2k}).  \label{ineq:xyz}
\end{align} 

Applying chain rules of entropy on $T_{\rm H}$ from the root to leaves gives the following decomposition of the source entropy \cite{gallager78}.
\begin{align}
H(X) = \sum_{k \in \mathcal{U}} \left( q_{2k-1} + q_{2k} \right) h \left( \frac{ q_{2k}}{q_{2k-1} + q_{2k}} \right).\label{source_entropy}
\end{align}
Thus, the redundancy of the optimal binary AIFV code, $r_{\rm AIFV}$, defined by \eqref{def:aifv_redundancy} is upper bounded from \eqref{ineq:xyz} and \eqref{source_entropy} as follows.
\begin{align}
r_{{\rm AIFV}} & \leq L_{T_0}\cdot P(T_0) + L_{T_1} \cdot P(T_1) - H(X)\nonumber\\
& =   \left[q_1 + q_2 - h \left(\frac{q_{2}}{q_{1} + q_{2}} \right) \right] \nonumber\\
& \ \ + \sum_{k \in \mathcal{K}} \left[2q_{2k} - (q_{2k} + q_{2k-1})h \left(\frac{q_{2k}}{q_{2k-1} + q_{2k}} \right) + q_2q_{2k-1} \right] \nonumber \\
& \ \ +\sum_{k \in \mathcal{U} \setminus (\mathcal{K} \cup \{ 1\})} (q_{2k-1} + q_{2k}) \left( 1 - h \left( \frac{q_{2k}}{q_{2k-1} + q_{2k}} \right) \right). \label{ha}
\end{align}%

%Note that in \eqref{ha}, we decompose the sum on $\mathcal{U}$ into three terms each of which is summed over $\{1\}, \ \mathcal{K}$, and $\mathcal{U} \setminus (\mathcal{K} \cup \{ 1\})$. First, suppose $k \in \mathcal{K}$. It follows from the definition of $\mathcal{K}$ that $2q_{2k} < q_{2k-1}$. Thus, we can apply Lemma \ref{lem-2} with $w_1 \coloneqq q_{2k-1},~ w_2 \coloneqq q_{2k}$ and $q \coloneqq q_2 \leq \frac{1}{2}$ to each $k \in \mathcal{K}$. Next, suppose $k \in \mathcal{U} \setminus (\mathcal{K} \cup \{ 1\})$. If $q_{2k-1}$ is a leaf, $q_{2k-1} \leq 2q_{2k}$ holds since $k \notin \mathcal{K}$. If $q_{2k-1}$ is not a leaf, then by Lemma \ref{lem-1}, $q_{2k-1} \leq 2q_{2k}$ holds. In either case, $q_{2k-1} \leq 2q_{2k}$ holds. Thus, we can apply Lemma \ref{lem-3} to each $k \in \mathcal{U} \setminus (\mathcal{K} \cup \{ 1\})$. 

We can apply Lemma \ref{lem-2} with $w_1 \coloneqq q_{2k-1},~ w_2 \coloneqq q_{2k}$ and $q \coloneqq q_2 \leq \frac{1}{2}$ to the second term of \eqref{ha} since $2q_{2k}<q_{2k-1}$ from the definition of $K$.
On the other hand, for $k \in \mathcal{U}\setminus(\mathcal{K}\cup\{1\})$,
if $q_{2k-1}$ is a leaf then $q_{2k-1}\leq2q_{2k}$ holds from the definition of $K$.
If $q_{2k-1}$ is not a leaf then $q_{2k-1}\leq2q_{2k}$ still holds from Lemma \ref{lem-1}.
Thus, we can apply Lemma \ref{lem-3} to the third term of \eqref{ha}.
Combining these with $q_1 + q_2 = 1$, we get
\begin{align}
r_{{\rm AIFV}} &\leq 1 - h(q_1) +   \sum_{k \in \mathcal{K}} q_2(q_{2k-1} - q_{2k})  + \sum_{k \in \mathcal{U} \setminus (\mathcal{K} \cup \{ 1\})} \frac{1}{4} (q_{2k-1} - q_{2k}) \nonumber \\
&\stackrel{({\rm a})}{\leq} 1 - h(q_1) + q_2\sum_{k = 2}^{K-1} (q_{2k-1} - q_{2k}) \nonumber\\
&\stackrel{({\rm b})}{<} 1 - h(q_1) + q_2q_3 \nonumber\\
& \stackrel{({\rm c})}{\leq} 1 - h(q_1) + q_2^2 \nonumber \\
& = q_1^2 - 2q_1 + 2 - h(q_1),
\end{align}
where (a) holds since $\frac{1}{4} < \frac{1}{3} \leq 1 - q_1 = q_2$, and (b) and (c) hold since the sequence $\{q_k\}$ is non-increasing. \qed
\end{myproof}

We now turn to the proof of Theorem \ref{thm-1}.

\begin{myproof} [Proof of Theorem \ref{thm-1}] 
First, consider the case of $\frac{1}{2} \leq p_{{\rm max}} \leq \frac{-1+\sqrt{5}}{2}$. Since $q_1 = p_{\max}$ for $p_{\max} \geq \frac{1}{2}$, it follows that $\frac{1}{2} \leq q_1 \leq \frac{-1+\sqrt{5}}{2} < \frac{2}{3}$.  Applying Lemma \ref{lem-4}, we get the upper bound on the redundancy as $p_{\max}^2 - 2p_{\max} + 2 - h(p_{\max})$.
\begin{spacing}{1.04}
Next, we prove the bound for $\frac{-1+\sqrt{5}}{2} \leq p_{\max} < 1$. The proof follows the same line as the proof of Lemma \ref{lem-4}. First, transform $T_{\rm base}$ into $T_0$ by the operation depicted in Fig.~\ref{fig:trans2} and also transform $T_{\rm base}$ into $T_1$ as illustrated in Fig.~\ref{fig:trans1}. 
\end{spacing}
\begin{figure}[t]
  \centering
  \includegraphics[width=6.5cm]{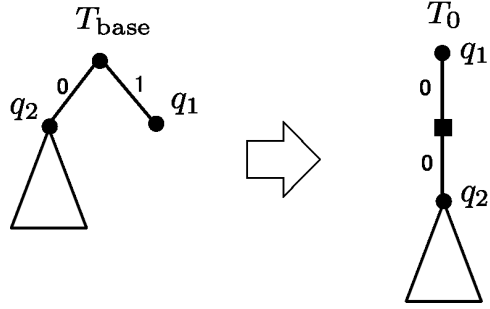}
  \caption{Transformation of $T_{\rm base}$ into $T_0$.}
  \label{fig:trans2}
\end{figure}
Then, $T_0$ and $T_1$ are valid binary AIFV code trees. In the same way as Lemma \ref{lem-4}, we can show that the stationary probabilities are given by $P(T_0) = \frac{1- \sum_{k \in \mathcal{K}}q_{2k-1}}{1 + q_1}$ and $P(T_1) = \frac{q_1 + \sum_{k \in \mathcal{K}}q_{2k-1}}{1 + q_1}$. As before, the redundancy of the optimal binary AIFV code can be upper bounded as follows.
\begin{align}
r_{{\rm AIFV}} & \leq L_{T_0}\cdot P(T_0) + L_{T_1} \cdot P(T_1) - H(X)\nonumber\\
& = \left[2q_2 + \frac{q_1^2}{1+q_1} - h\left(\frac{q_{2}}{q_{1} + q_{2}} \right)\right] \nonumber\\
&\ \ + \sum_{k \in \mathcal{K}} \left[ 2q_{2k} - (q_{2k} + q_{2k-1})h \left(\frac{q_{2k}}{q_{2k-1} + q_{2k}} \right) + \frac{q_1q_{2k-1}}{1+q_1} \right]\nonumber\\
& \ \ +  \sum_{k \in \mathcal{U} \setminus (\mathcal{K} \cup \{ 1\})} (q_{2k-1} + q_{2k}) \left( 1 - h \left( \frac{q_{2k}}{q_{2k-1} + q_{2k}} \right) \right). \label{eq:22x}
\end{align}

We can apply Lemma \ref{lem-2} with $w_1 \coloneqq q_{2k-1},~w_2 \coloneqq q_{2k}$ and $q \coloneqq \frac{q_1}{1+q_1} \leq \frac{1}{2}$ to the second term of \eqref{eq:22x}. Also, we can apply Lemma \ref{lem-3} to the third term of \eqref{eq:22x}. Combining these facts with $q_1 + q_2 = 1$ and $q_1 = p_{\max}$, we get
\begin{align}
r_{{\rm AIFV}} &\leq \left[2(1-q_1) + \frac{q_1^2}{1 + q_1}- h(q_1)\right] \nonumber \\
&\ \  + \sum_{k \in \mathcal{K}} \frac{q_1}{1 + q_1}(q_{2k-1} - q_{2k}) + \sum_{k \in \mathcal{U} \setminus (\mathcal{K} \cup \{ 1\})} \frac{1}{4} (q_{2k-1} - q_{2k}) \nonumber \\
&\stackrel{({\rm d})}{\leq} 2(1-q_1) + \frac{q_1^2}{1 + q_1}- h(q_1)+  \frac{q_1}{1+q_1}\sum_{k = 2}^{K-1} (q_{2k-1} - q_{2k}) \nonumber \\
%& \leq 2(1-q_1) + \frac{q_1^2}{1 + q_1}- h(q_1) +  \frac{q_1}{1+q_1}q_3 \label{eq:32}\\
&\stackrel{({\rm e})}{<} 2(1-q_1) + \frac{q_1^2}{1 + q_1}- h(q_1)+  \frac{q_1q_2}{1+q_1} \nonumber\\
& = \frac{-2p_{{\rm max}}^2+p_{{\rm max}}+2}{1+p_{{\rm max}}} - h(p_{{\rm max}}),
\end{align}%
where (d) holds since $\frac{1}{4} < \frac{3-\sqrt{5}}{2} \leq \frac{q_1}{1 + q_1}$, and (e) holds since the sequence $\{q_k\}$ is non-increasing. 

\textcolor{black}{To prove that the derived bound is tight and coincides with the worst-case redundancy for sources with $p_{\rm max}$}, it is sufficient to show that there exists a source for every $p_{\max} \geq \frac{1}{2}$ such that the \textcolor{black}{redundancy of the optimal code for the source} attains the bound arbitrarily closely. In particular, we show that the \textcolor{black}{redundancy of the optimal code} for a source with probabilities $(p_{\max}, 1 - p_{\max} - \delta, \delta)$ satisfies the bound with equality in the limit of $\delta \rightarrow 0$. Note that for $|\mathcal{X}|=3$, there exist only four possible tree structures for each code tree, $T_0$ and $T_1$.  By examining all the possible combinations of the structures, it can be shown that the optimal binary AIFV codes are as illustrated in Fig.~\ref{fig_sim} for each range of $p_{\max}$. We see that the redundancy of the codes coincides with the bound in the limit of $\delta \rightarrow 0$. \qed
\begin{figure}[t]
\centering
\subfloat[$\frac{1}{2} \leq p_{{\rm max}} \leq \frac{\sqrt{5}-1}{2}$.]{\includegraphics[width=5.0cm]{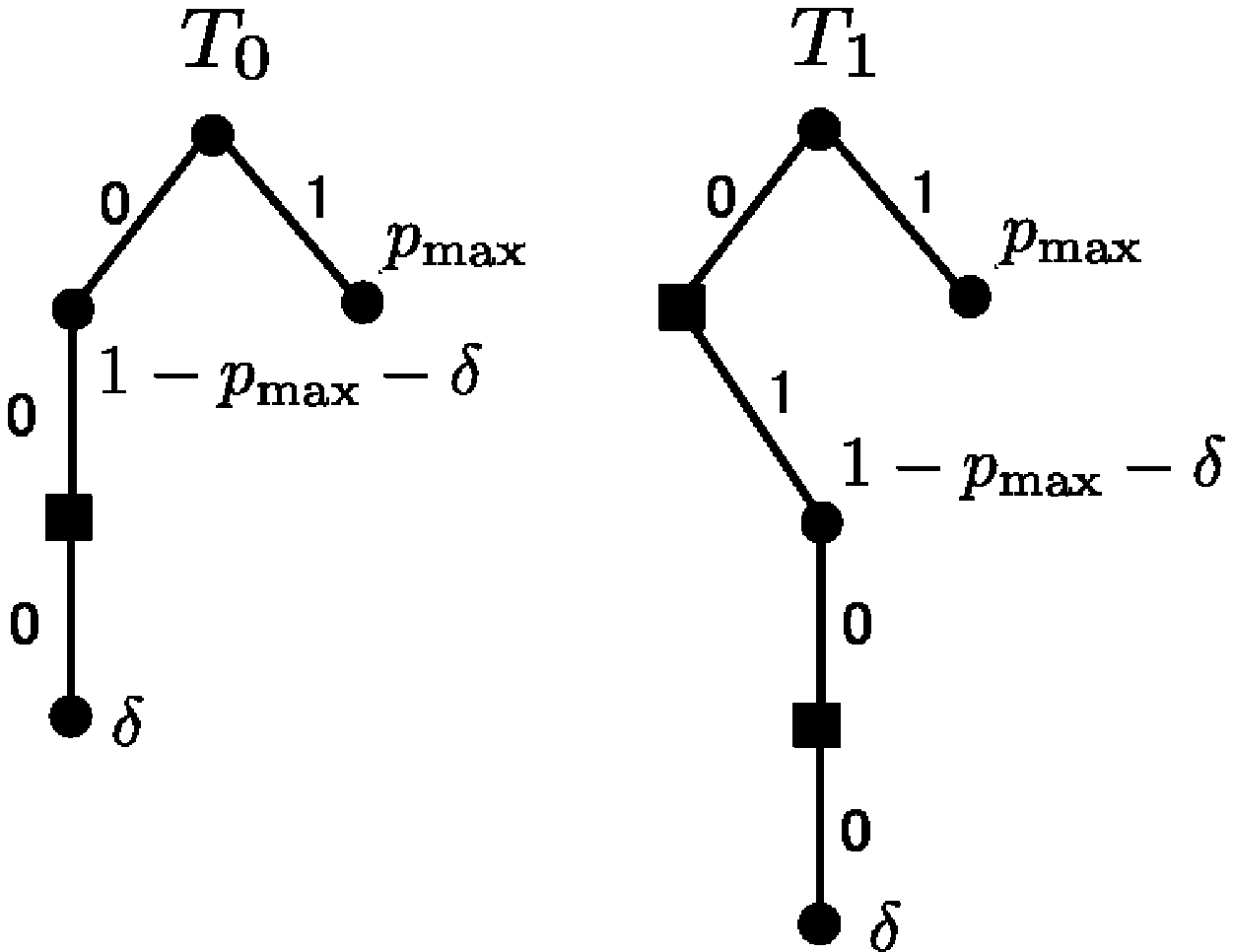}
\label{fig_first_case}}
\hfil
\subfloat[$\frac{\sqrt{5}-1}{2} \leq p_{{\rm max}} \leq 1$.]{\includegraphics[width=4.3cm]{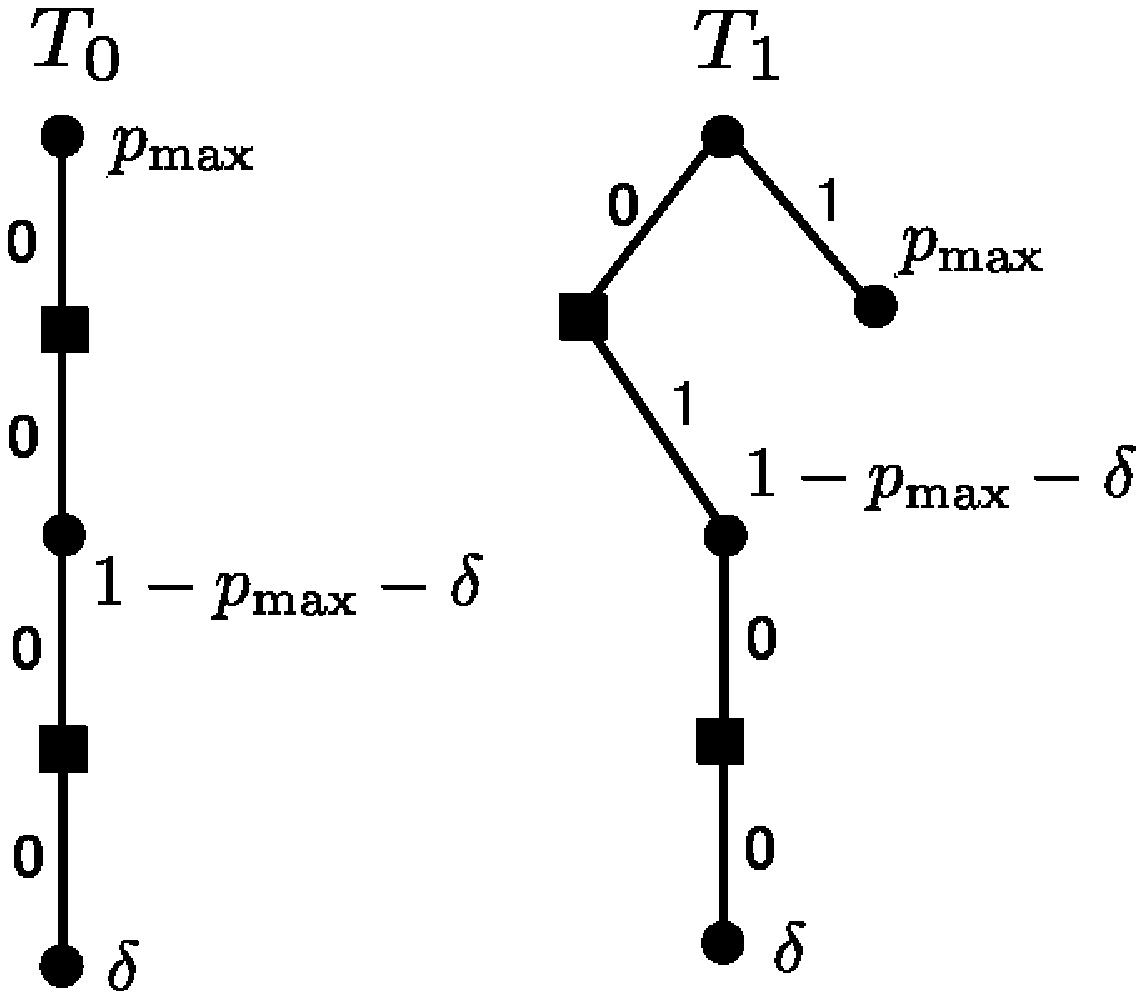}
\label{fig_second_case}}
\caption{\textcolor{black}{The optimal binary AIFV code trees for the worst-case source with $p_{\rm max}$.}}
\label{fig_sim}
\end{figure}
\end{myproof}

We proceed to prove Theorem \ref{thm-2}, which provides a simple bound on the redundancy of optimal binary AIFV codes for $p_{\max} < \frac{1}{2}$.

\begin{myproof}[Proof of Theorem \ref{thm-2}]
In the case of $p_{\max} < \frac{1}{2}$, we note that $|\mathcal{X}|\geq 3$ and thus, $q_1$ must be an internal node. It follows from Lemma \ref{lem-1} that $q_1 \leq 2q_2.$ Since $q_1 + q_2 = 1$, we get $\frac{1}{2} \leq q_1 \leq \frac{2}{3}$. By Lemma \ref{lem-4}, we obtain
\begin{align}
r_{{\rm AIFV}} \leq \max_{\frac{1}{2} \leq q_1 \leq \frac{2}{3}} q_1^2 - 2q_1 + 2 - h(q_1) = \frac{1}{4}. 
\end{align}
\qed
\end{myproof}

We turn to the proof of Theorem \ref{lower_bb}, which lower bounds the worst-case redundancy of optimal binary AIFV-$m$ codes.

\begin{myproof}[Proof of Theorem \ref{lower_bb}]
Consider a source $a \in \mathcal{X}$ with $P(a) = 1 - \delta$, where $\delta$ is arbitrarily close to 0. Our goal is to show that the redundancy of optimal binary AIFV-$m$ codes for the source must be larger than or equal to $\frac{1}{m}$ in order for the code to be uniquely decodable.

Let $G = (V, E)$ be a directed graph,
where
\begin{align}
V&\equiv\{i\in\{ \textcolor{black}{0,1,\cdots,m-1}\}| \mbox{Encoding procedure is able to reach $T_i$ from $T_0$}\},\nonumber\\
E&\equiv\{(i,j)| \mbox{
%the code tree moves to $T_j$ when `$a$' is encoded by $T_i$
$T_j$ is used after encoding of `$a$' by $T_i$}\}.
\end{align}
By a slight abuse of notation we sometimes identify an index $i\in V$ with the code tree $T_i$ in the following.
Here note that $G$ contains disjoint loops\footnote{\textcolor{black}{The definition of the loop includes a self-loop.}} because the number of outgoing edges is exactly one for each node.
Let $S_1, S_2, \cdots, S_k\subset V$ be sets of nodes in each loop,
where
$k$ is the number of loops, and
let $S_0\equiv V\setminus (S_1 \cup S_2 \cup \cdots \cup S_k)$.
%Here each $S_i,\,i\ge 1,$ forms a cycle since the number of outgoing edges is exactly one.

%Now we consider the Markov process over code trees $\{T_i: i\in V\}$.
Now we consider the Markov chain corresponding to the code tree of a binary AIFV-$m$ code.
The transition probability from code tree $T_i$ to $T_j$ is denoted by
$P_{\delta}(T_j|T_i)$.
The stationary probability of the code tree $T$ is denoted by $P_{\delta} (T)$ and
we define $P_{\delta}(S) = \sum_{i \in S} P_{\delta}(T_i)$ for $S \subset V$.

Let $P_{\delta}^{(l)}(T_j|T_i)$ be
the transition probability for $l$ transitions,
that is,
\begin{align}
P_{\delta}^{(l)}(T_j|T_i)=
\begin{cases}
P_{\delta}(T_j|T_i),&l=1,\\
\sum_{a\in V} P_{\delta}^{(l-1)}(T_a|T_i)P_{\delta}(T_j|T_a),&l\ge 2.\\
\end{cases}
\end{align}
Here note that the code tree is not in $S_0$ after encoding of $\underbrace{aa\cdots a}_{m}$ from any code tree
since $S_0$ contains no loop and $|S_0|\le m$. Thus, for all $j \in S_0$, we have
\textcolor{black}{
\begin{align}
\sum_{i \in S_0}P_{\delta}^{(m)}(T_i | T_j)  &= 1 - \sum_{i \in V \setminus S_0}P_{\delta}^{(m)}(T_i | T_j) \nonumber \\
& \leq 1 - (1 - \delta)^m.
\end{align}}
Therefore, we have
\begin{align}
P_{\delta}(S_0)
&=
\sum_{i\in S_0}P_{\delta}(T_i)\nonumber\\
&=
\sum_{j\in V}\sum_{i\in S_0}P_{\delta}(T_j)P_{\delta}^{(m)}(T_i|T_j)\nonumber \\
&\le
\sum_{j\in V}P_{\delta}(T_j)(1-(1-\delta)^m)\nonumber \\
&=
1-(1-\delta)^m.
\end{align}

Hence, we get 
\begin{align}
P_{\delta}\left(\bigcup_{i=1}^k S_i\right)
\geq (1-\delta)^m. \label{lower9}
\end{align}

Consider the set of code trees $S_i$, $i\ge 1$.
We write its elements by $S_i=\{j_1, j_2, \dots, j_s\}$
so that the loop of $S_i$ in $G$ is represented by $v_{j_1} \to v_{j_2} \to \cdots \to v_{j_s} \to v_{j_1}$.
Recall that each transition on the graph corresponds to encoding of a source symbol `$a$'. Therefore, we get
\begin{align}
P_{\delta}(T_{j_2}) &\geq (1-\delta)P_{\delta}(T_{j_1}) \nonumber\\
P_{\delta}(T_{j_3}) &\geq (1-\delta)P_{\delta}(T_{j_2}) \nonumber \\
&\ \vdots \nonumber \\
P_{\delta}(T_{j_s}) &\geq (1-\delta)P_{\delta}(T_{j_{s-1}}) \nonumber \\
P_{\delta}(T_{j_1}) &\geq (1-\delta)P_{\delta}(T_{j_{s}}). 
\end{align}
%for $2 \leq n \leq j_s$,
%\begin{align}
%P_{\delta}(T_{j_n}) \geq (1-\delta) P_{\delta}(T_{j_{n-1}})
%\end{align}
Combining these inequalities, we obtain
\begin{align}
P_{\delta}(T_{j_n}) \leq \frac{P_{\delta}(T_{j_{n+1}})}{1- \delta}
\le \cdots
 \leq \frac{P_{\delta}(T_{j_s})}{(1- \delta)^{s-n}} .
\end{align}
Hence,
\begin{align}
P_{\delta}(S_i) &= \sum_{n = 1}^{s} P_{\delta}(T_{j_n})\nonumber \\
& \leq \sum_{n = 1}^{s} \frac{P_{\delta}(T_{j_s})}{(1-\delta)^{s-n}}\nonumber \\
& = P_{\delta}(T_{j_s}) \cdot \sum_{n = 1}^{s} (1-\delta)^{-(s-n)},
\end{align}
which implies
\begin{align}
P_{\delta}(T_{j_s})
\geq \frac{P_{\delta}(S_i)}{\sum_{n = 1}^{s} (1-\delta)^{n-s}}
\geq \frac{P_{\delta}(S_i)}{\sum_{n = 1}^{m} (1-\delta)^{n-m}}.
\end{align}
Similarly, we can bound $P_{\delta}(T_{j_n}),\,1 \le n \le s$, as
\begin{align}
P_{\delta}(T_{j_n}) \geq \frac{P_{\delta}(S_i)}{\sum_{n^{\prime} = 1}^{m} (1-\delta)^{n^{\prime}-m}}. \label{good2}
\end{align}

Here note that at least one of $T_{j_1}, T_{j_2}, \cdots ,T_{j_s}$ must assign `$a$' to its non-root node in order for the code to be uniquely decodable. Otherwise, $\underbrace{a\cdots a}_s$ is encoded to null codeword with length 0, and hence,
source sequences $x\underbrace{aa\cdots a}_{s}x^{\prime}$ and $xx^{\prime}$ are encoded to exactly the same codeword sequence.
In such a case, the code is no longer uniquely decodable. Therefore, $\sum_{n = 1}^{s} L_{T_{j_n}} \geq 1 - \delta$ holds. Combining this with \eqref{good2}, we obtain a lower bound of the average code length as follows.
\begin{align}
\sum_{j\in V}P_{\delta}(T_j)L_{T_j}
&\ge
\sum_{i=1}^k\sum_{j \in S_i} P_{\delta}(T_j)L_{T_j}\nonumber\\
&\ge
\sum_{i=1}^k\sum_{j \in S_i} \frac{P_{\delta}(S_i)}{\sum_{n^{\prime} = 1}^{m} (1-\delta)^{n^{\prime}-m}}
L_{T_j}\nonumber\\
&\ge
\sum_{i=1}^k\frac{(1-\delta)P_{\delta}(S_i)}{\sum_{n^{\prime} = 1}^{m} (1-\delta)^{n^{\prime}-m}}\nonumber\\%\label{lower11}\\
&\ge
\frac{(1-\delta)^{m+1}}{\sum_{n^{\prime} = 1}^{m} (1-\delta)^{n^{\prime}-m}},\label{lower13}
\end{align}
where the last inequality follows from \eqref{lower9}. 

Finally, the redundancy of the binary AIFV-$m$ codes for $\delta \to 0$
is lower bounded by
\begin{align}
\liminf_{\delta \to 0} \left( \sum_{i \in V} P_{\delta}(T) L_{T_i} - H_{\delta}(X)\right)
&\geq \liminf_{\delta \to 0} \left( \frac{(1-\delta)^{m+1}}{ \sum_{n'=1}^{m}(1- \delta)^{n'-m}} - H_{\delta}(X) \right) \nonumber \\
&= \frac{1}{m},
\end{align}
where $H_{\delta}(X)$ is the source entropy.
This argument holds for any AIFV-$m$ code and we get Theorem \ref{lower_bb}.
\qed
\end{myproof}

We then turn to the proofs of Theorems \ref{bound3} and \ref{bound4}, 
%which provide tight upper bounds on the redundancy of optimal binary AIFV-$3$ codes and AIFV-$4$ codes, respectively.
\textcolor{black}{which provide the worst-case redundancy of optimal binary AIFV-$3$ codes and AIFV-$4$ codes, respectively.} 

\begin{myproof}[Proof of Theorem \ref{bound3}]
\textcolor{black}{By Theorem \ref{lower_bb}, the worst-case redundancy of optimal binary AIFV-3 codes is larger than or equal to $\frac{1}{3}$. Therefore, to show that the worst-case redundancy is exactly $\frac{1}{3},$ it is sufficient to show that there exist binary AIFV-3 codes whose redundancy is at most $\frac{1}{3}$.}

Consider the case of $p_{\max} \geq \frac{1}{2}$. Then, $q_1$ is a leaf node. We transform a Huffman code tree $T_{\rm H}$ into binary AIFV-3 code trees by the transformation shown in Fig.~\ref{fig:aifv3}. It is easy to see that these code trees satisfy the conditions of binary AIFV-3 codes. The transition matrix $R$ is calculated as
\begin{align}
R = 
\begin{pmatrix}
1-p_{\max} & 0 & p_{\max} \\
1 & \textcolor{black}{0} & 0 \\
1-p_{\max} & p_{\max} & 0
\end{pmatrix}.
\end{align}
Thus, the stationary probabilities are given by 
\begin{align}
P(T_0) &= \frac{1}{1 + p_{\max} +  p_{\max}^2},\label{stationary3from} \\ 
P(T_1) &= \frac{p_{\max}^2}{1 + p_{\max} +  p_{\max}^2},\\
P(T_2) &= \frac{p_{\max}}{1 + p_{\max} +  p_{\max}^2}. \label{stationary3to}
\end{align}

Let $r_{T_i}~i = 1,2,3$ denote the redundancy of an FV code with the code tree $T_i$. Applying Theorem \ref{huffman_bound} gives upper bounds on $r_{T_i}$ as follows.
\begin{align}
r_{T_0} &< 2 - p_{\max} - h({p_{\max}}) - q_1 + 2q_2,\label{r3from} \\
r_{T_1} &<  2 - p_{\max} - h({p_{\max}}) +q_2, \\
r_{T_2} &< 2 - p_{\max} - h({p_{\max}})  - q_1 + 2q_2. \label{r3to}
\end{align}

%Let $r_{3}$ be the redundancy of binary AIFV-3 codes shown in the right side of Fig.~\ref{fig:aifv3}. 
Let $r_{{\rm AIFV}\text{-}m}(p_{\max})$ be the redundancy of optimal binary AIFV-$m$ codes for a source with $p_{\max}$. Combining \eqref{stationary3from}--\eqref{stationary3to} with \eqref{r3from}--\eqref{r3to}, $r_{{\rm AIFV}\text{-}3}(p_{\max})$ is upper bounded as follows.
\begin{align}
r_{{\rm AIFV}\text{-}3}(p_{\max}) &\leq \sum_{i = 0}^2 P(T_i)\cdot r_{T_i} \nonumber \\
&< 2 - p_{\max} - h(p_{\max}) + \frac{1+p_{\max}}{1+p_{\max} + p_{\max}^2} \left(-p_{\max} + 2(1-p_{\max}) \right)\nonumber\\
& \qquad +  \frac{p_{\max}^2}{1+p_{\max} + p_{\max}^2} (1-p_{\max}) \nonumber \\
& \equiv f_3(p_{\max}). \label{eq:ww}
\end{align}

Note that our definition of binary AIFV-3 codes includes the original binary AIFV codes. Therefore, for $p_{\max} \geq \frac{1}{2}$, $r_{{\rm AIFV}\text{-}3}(p_{\max})$ can be upper bounded as follows.
\begin{align}
r_{{\rm AIFV}\text{-}3}(p_{\max}) \leq \min \{ f({p_{\max}}),\  f_3(p_{\max})\}, 
\end{align}
where the functions $f(\cdot)$ and $f_3(\cdot)$ are defined in Theorem \ref{thm-1} and \eqref{eq:ww}, respectively. Fig.~\ref{fig:aifv3_bound} illustrates the function $\min \{ f({p_{\max}}),\  f_3(p_{\max})\}$. We see that $r_{{\rm AIFV}\text{-}3}(p_{\max}) \leq \min \{ f({p_{\max}}),\  f_3(p_{\max})\} < \frac{1}{3}$ holds for every $p_{\max} \geq \frac{1}{2}$. Also, Theorem \ref{thm-2} implies that for $p_{\max} < \frac{1}{2}$, $r_{{\rm AIFV}\text{-}3}(p_{\max}) < \frac{1}{4}$ holds. In summary, $r_{{\rm AIFV}\text{-}3}(p_{\max}) < \frac{1}{3}$ holds for every $p_{\max} \in (0,1)$. Hence, the redundancy of optimal binary AIFV-3 codes is upper bounded as
\begin{align} 
r_{{\rm AIFV\text{-}3}} \leq {\sup_{p_{\max} \textcolor{black}{\in (0,1)}}}\  r_{{\rm AIFV}\text{-}3}(p_{\max}) = \frac{1}{3}.
\end{align} 
\qed
\begin{figure}[t]
  \centering
  \includegraphics[width=9cm]{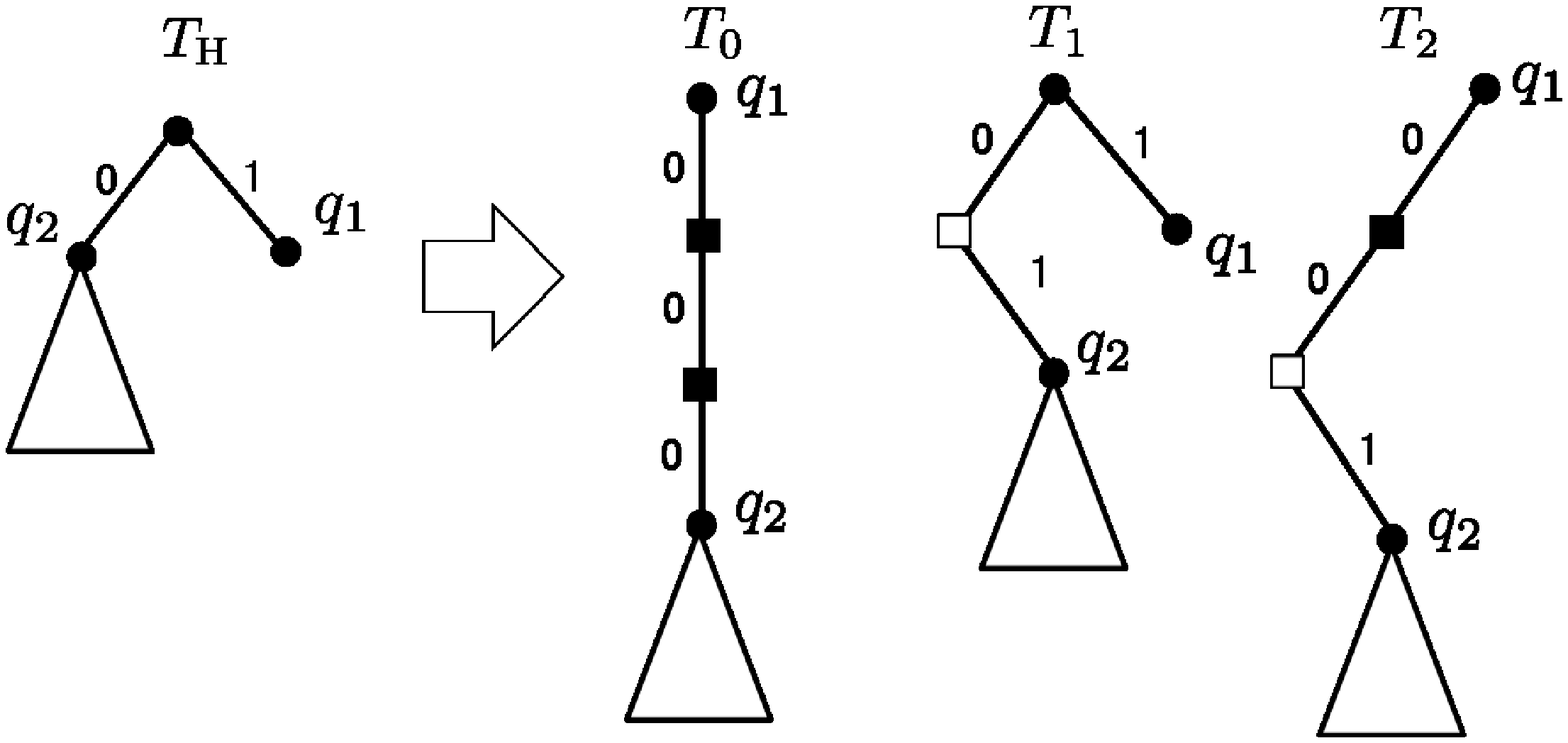}
  \caption{Transformation from $T_{\rm H}$ to binary AIFV-3 code trees.}
  \label{fig:aifv3}
  \vspace{+1cm}
\end{figure}

\begin{figure}[t]
  \centering
  \includegraphics[width=8.5cm]{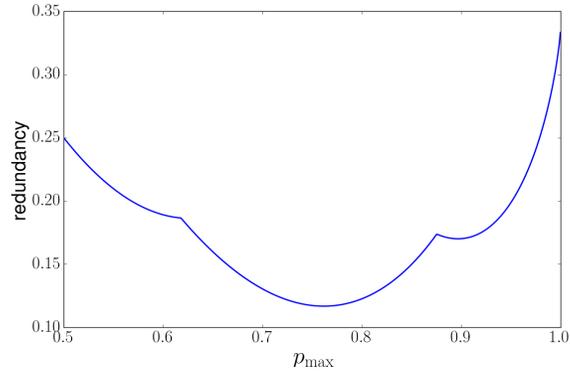}
  \caption{The redundancy upper bounds of optimal binary AIFV-3 codes in terms of $p_{\max}$.}
  \label{fig:aifv3_bound}
\end{figure}
\end{myproof}

\begin{myproof}[Proof of Theorem \ref{bound4}]
The proof follows the same line as the previous proof of binary AIFV-3 codes. 
\textcolor{black}{By Theorem \ref{lower_bb}, the worst-case redundancy of optimal binary AIFV-4 codes is larger than or equal to $\frac{1}{4}$. Therefore, to show that the worst-case redundancy is exactly $\frac{1}{4},$ it is sufficient to show that there exist binary AIFV-4 codes whose redundancy is at most $\frac{1}{4}$.}

%By Theorem \ref{lower_bb}, a tight upper bound on the redundancy of optimal binary AIFV-4 codes is larger than or equal to $\frac{1}{4}$. Therefore, to show that the tight upper bound is exactly $\frac{1}{4},$ it is sufficient to show that the upper bound is at most $\frac{1}{4}$. 

Consider the case of $p_{\max} \geq \frac{1}{2}$. Then, $q_1$ is a leaf node. We transform a Huffman code tree $T_{\rm H}$ into binary AIFV-4 code trees by the transformation shown in Fig.~\ref{fig:aifv4}. It is easy to see that these code trees satisfy the conditions of binary AIFV-4 codes. The transition matrix $R$ is calculated as
\begin{align}
R = 
\begin{pmatrix}
1-p_{\max} & 0 & 0 & p_{\max} \\
1 & 0 & 0 & 0 \\
1-p_{\max} & p_{\max} & 0 & 0 \\
1-p_{\max} & 0 & p_{\max} & 0 
\end{pmatrix}.
\end{align}
Thus, the stationary probabilities are given by
\begin{align}
P(T_0) &= \frac{1}{1 + p_{\max} +  p_{\max}^2 + p_{\max}^3}, \label{stationary4from}\\
P(T_1) &= \frac{p_{\max}^3}{1 + p_{\max} +  p_{\max}^2 + p_{\max}^3}, \\   
P(T_2) &= \frac{p_{\max}^2}{1 + p_{\max} +  p_{\max}^2 + p_{\max}^3},\\
P(T_3) &= \frac{p_{\max}}{1 + p_{\max} +  p_{\max}^2 + p_{\max}^3}. \label{stationary4to}
\end{align}

Let $r_{T_i}~1 \leq i \leq 4$ denote the redundancy of an FV code with the code tree $T_i$. Applying Theorem \ref{huffman_bound} gives upper bounds on $r_{T_i}$ as follows.
\begin{align}
r_{T_0} &< 2 - p_{\max} - h({p_{\max}}) - q_1 + 3q_2, \label{r4from} \\
r_{T_1} &<  2 - p_{\max} - h({p_{\max}}) + q_2, \\
r_{T_2} &<  2 - p_{\max} - h({p_{\max}}) - q_1 + 2q_2,\\
r_{T_3} &< 2 - p_{\max} - h({p_{\max}})  - q_1 + 3q_2. \label{r4to}
\end{align}

Combining \eqref{stationary4from}--\eqref{stationary4to} with \eqref{r4from}--\eqref{r4to}, $r_{{\rm AIFV}\text{-}4}(p_{\max})$ is upper bounded as follows.
\begin{align}
r_{{\rm AIFV}\text{-}4}(p_{\max}) &\leq \sum_{i = 0}^3 P(T_i)\cdot r_{T_i} \nonumber \\
&< 2 - p_{\max} - h(p_{\max}) + \frac{1+p_{\max}}{1+p_{\max} + p_{\max}^2 + p_{\max}^3}  \left(-p_{\max} + 3(1-p_{\max}) \right)\nonumber \\
& \qquad +  \frac{p_{\max}^2}{1+p_{\max} + p_{\max}^2 + p_{\max}^3} (-p_{\max} + 2(1-p_{\max})) \nonumber \\
&\qquad +   \frac{p_{\max}^3}{1+p_{\max} + p_{\max}^2 + p_{\max}^3} (1-p_{\max}) \nonumber \\
& \equiv f_4(p_{\max}). \label{eq:www}
\end{align}
Thus, for $p_{\max} \geq \frac{1}{2}$, $r_{{\rm AIFV}\text{-}4}(p_{\max})$ is upper bounded as
\begin{align}
r_{{\rm AIFV}\text{-}4}(p_{\max}) \leq \min \{ f({p_{\max}}),\  f_4(p_{\max})\}. 
\end{align}
Fig.~\ref{fig:aifv4_bound} illustrates the function $\min \{ f({p_{\max}}),\  f_4(p_{\max})\}$. We see that $r_{{\rm AIFV}\text{-}4}(p_{\max}) \leq \min \{ f({p_{\max}}),\  f_4(p_{\max})\} < \frac{1}{4}$ holds for every $p_{\max} \geq \frac{1}{2}$. Also, Theorem \ref{thm-2} implies that for $p_{\max} < \frac{1}{2}$, $r_{{\rm AIFV}\text{-}4}(p_{\max}) < \frac{1}{4}$ holds. In summary, $r_{{\rm AIFV}\text{-}4}(p_{\max}) < \frac{1}{4}$ holds for every $p_{\max} \in (0,1)$. Hence, the redundancy of optimal binary AIFV-4 codes is upper bounded as
\begin{align} 
r_{{\rm AIFV\text{-}4}} \leq {\sup_{p_{\max} \textcolor{black}{\in (0,1)}}} \ r_{{\rm AIFV}\text{-}4}(p_{\max}) = \frac{1}{4}.
\end{align} 
\qed

\begin{figure}[t]
  \centering
  \includegraphics[width=10cm]{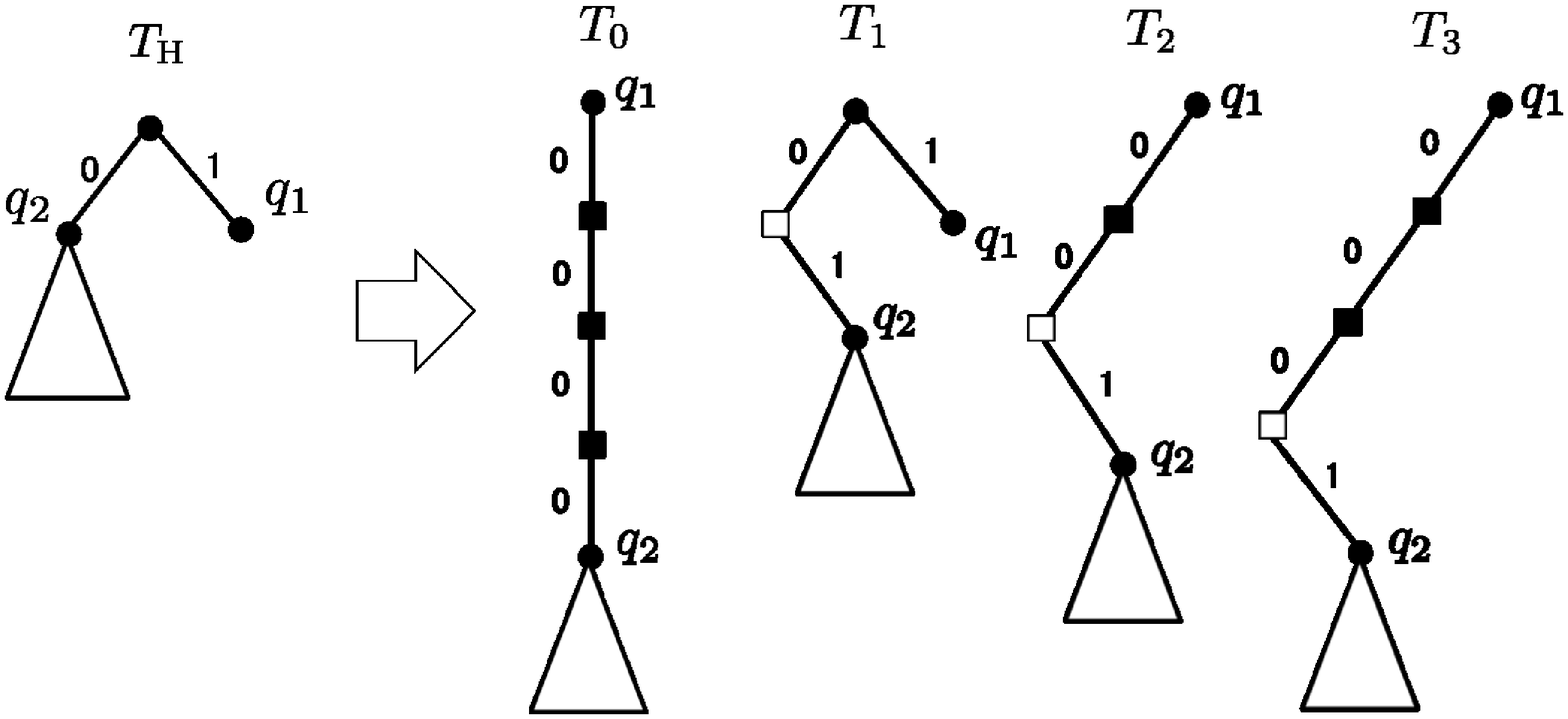}
  \caption{Transformation from $T_{\rm H}$ to binary AIFV-4 code trees.}
  \label{fig:aifv4}
    \vspace{+1cm}
\end{figure}

\begin{figure}[t]
  \centering
  \includegraphics[width=8.5cm]{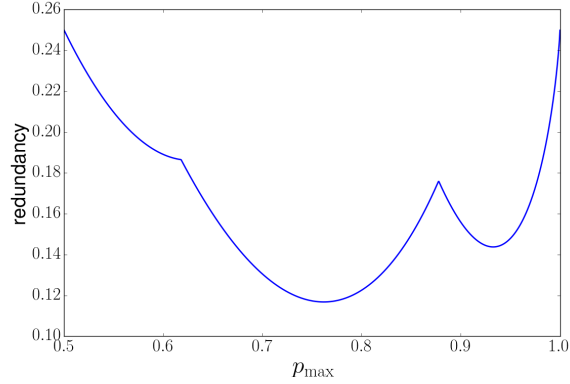}
  \caption{The redundancy upper bounds of binary AIFV-4 codes in terms of $p_{\max}$.}
  \label{fig:aifv4_bound}
\end{figure}
\end{myproof}

We then provide a simple proof of Theorem \ref{pmax_one}, which justifies our conjecture that the redundancy of binary AIFV-$m$ codes is upper bounded by $\frac{1}{m}$ for any source.

\begin{myproof}[Proof of Theorem \ref{pmax_one}]
\textcolor{black}{In the proof of Theorem \ref{lower_bb}, we see that the redundancy of optimal binary AIFV-$m$ codes is larger than or equal to $\frac{1}{m}$ for $p_{\max} \rightarrow 1.$ Hence, to show that the redundancy is $\frac{1}{m}$ as $p_{\max} \rightarrow 1$, it is sufficient to show that there exists a binary AIFV-$m$ code such that its redundancy is $\frac{1}{m}$ as $p_{\max} \rightarrow 1.$}

Suppose $p_{\max} \geq \frac{1}{2}$. Consider binary AIFV-$m$ code trees obtained by the transformation shown in Fig.~\ref{fig:aifvm_bound}, which is a generalization of Figs.~\ref{fig:aifv3} and \ref{fig:aifv4} to the case of binary AIFV-$m$ codes. Analogous to \eqref{stationary3from}--\eqref{stationary3to} and \eqref{stationary4from}--\eqref{stationary4to}, it can be shown that the resulting binary AIFV-$m$ code trees, denoted by $T_0, T_1, \cdots, T_{m-1}$, have the stationary probabilities as
\begin{figure}[t]
  \centering
  \includegraphics[width=11.5cm]{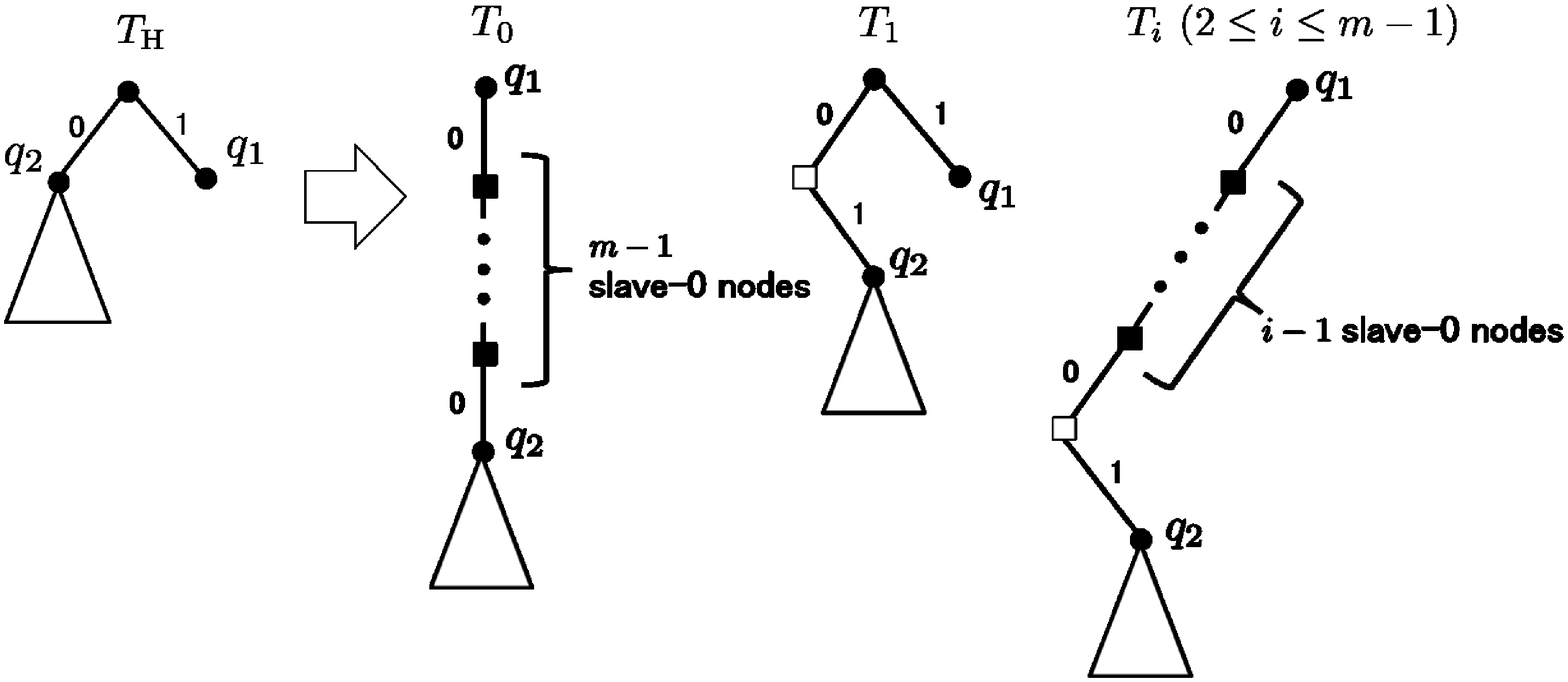}
  \caption{\textcolor{black}{Transformation from $T_{\rm H}$ to binary AIFV-$m$ code trees.}}
  \label{fig:aifvm_bound}
\end{figure}
\begin{align} \label{transition}
P(T_i) = \left\{ \begin{array}{ll}
\frac{1}{\sum_{k = 0}^{m-1} p_{\max}^k} & \ \  \text{if $i=0$,} \\
\frac{p_{\max}^2}{\sum_{k = 0}^{m-1} p_{\max}^k}  & \ \  \text{if $i=1$,}\\
\frac{p_{\max}^{m-i}}{\sum_{k = 0}^{m-1} p_{\max}^k}  & \ \  \text{if $2 \leq i \leq m-1$.}\\
\end{array} \right.
\end{align}
In the limit of $p_{\max} \to 1$, stationary probabilities for all the code trees become $\frac{1}{m}$. In addition, we see from Fig.~\ref{fig:aifvm_bound} that the average code length tends to 0 for $T_{i},\ i \neq 1$, and tends to 1 for $T_1.$ Hence, the average code length of the binary AIFV-$m$ code tends to $\frac{1}{m}$ as $p_{\max} \rightarrow 1.$ Since source entropy tends to 0 as $p_{\max} \rightarrow 1$, the redundancy of the binary AIFV-$m$ code tends to $\frac{1}{m}$ in the limit of $p_{\max} \rightarrow 1.$ Therefore, the redundancy of the binary AIFV-$m$ code in Fig.~\ref{fig:aifvm_bound} is $\frac{1}{m}$ in the limit of $p_{\max} \rightarrow 1.$ 
\qed
\end{myproof}
 
\textcolor{black}{Finally, we remark on the technical challenge in evaluating the worst-case redundancy of optimal binary AIFV-$m$ for $m \geq 5$. We first note that our analyses of the worst-case redundancy of optimal binary AIFV-$m$ codes for $m = 3$ and 4 rely on the fact that the redundancy upper bounds  shown in Fig.~\ref{fig:comparison} and Theorem \ref{thm-2} are below $\frac{1}{4}$ for all $p_{\rm max}$ that satisfy $0 < p_{\rm max} \leq 0.92$. Hence, when evaluating the worst-case redundancy of optimal binary AIFV-$m$ codes with $m = 3$ (resp. 4), we only need to ensure that the worst-case redundancy is small enough, i.e., below $\frac{1}{3}$ $\left(\text{resp. } \frac{1}{4} \right)$ for very large $p_{\rm max}$, i.e., $0.92 \leq p_{\rm max} < 1$. 
On the other hand, when showing that the worst-case redundancy of optimal binary AIFV-$m$ codes is $\frac{1}{m}$ for $m \geq 5$, we need to improve the bounds of Theorems \ref{thm-1} and \ref{thm-2} for most $p_{\rm max}$, including all $p_{\rm max}$ that satisfy $0 < p_{\rm max} < \frac{1}{2}$. This is because, when $m \geq 5$, the bounds provided by the theorems exceed $\frac{1}{m}$ for most $p_{\rm max}$. This makes it non-trivial for the redundancy analyses for the case of $m \leq 4$ to be extended to the case of $m \geq 5$.}

\section{Conclusion} \label{conclusion}
In this paper, we considered binary AIFV codes that use two code trees and decoding delay is at most two bits. 
\textcolor{black}{We showed that the worst-case redundancy of the optimal codes is $\frac{1}{2}.$}
We also extended the original binary AIFV codes \textcolor{black}{by allowing} the use of more code trees. We showed that when three and four code trees are allowed to be used, 
\textcolor{black}{the worst-case redundancy} of binary AIFV-3 codes and binary AIFV-4 codes are improved to $\frac{1}{3}$ and $\frac{1}{4}$, respectively. For the future research, it is interesting to derive \textcolor{black}{the worst-case redundancy of optimal binary AIFV codes in terms of $p_{\rm max}$ for $p_{\max}< \frac{1}{2}$, and compare it to its Huffman counterpart.} It may also be interesting to obtain other bounds (e.g., asymptotic redundancy \cite{szpankowski2000}) on redundancy of optimal binary AIFV codes, whose Huffman counterparts are already known. As regards binary AIFV-$m$ codes, it is an open problem to derive \textcolor{black}{the worst-case redundancy} of the \textcolor{black}{optimal} codes for any natural number $m$. Our conjecture is that under certain conditions on the size of alphabet, \textcolor{black}{the worst-case redundancy of optimal binary AIFV-$m$ codes is $\frac{1}{m}$.} Furthermore, it is necessary to explore efficient algorithms to obtain optimal (or near-optimal) codes in the class of binary AIFV-$m$ codes and \textcolor{black}{empirically compare their performance with Huffman codes for $\mathcal{X}^m$.}

%\section*{Acknowledgment}

\ifCLASSOPTIONcaptionsoff
  \newpage
\fi



%\bibliographystyle{IEEEtran}
%\begin{thebibliography}{1}
  %0.5em minus 0.4em\relax Harlow, England: Addison-Wesley, 1999.
%\end{thebibliography}
\bibliographystyle{IEEEtran}
\bibliography{reference}


%\newpage



%\vfill


\end{document}